\documentclass[letterpaper,twocolumn,screen,10pt]{article}
\usepackage{usenix}

\usepackage{xspace} 
\usepackage{enumitem}
\usepackage{color}
\usepackage{colortbl}
\usepackage{xcolor}
\usepackage{pifont}

\usepackage{epstopdf}
\usepackage{graphicx}
\usepackage[labelfont=bf, textfont=bf]{caption}
\usepackage{subcaption}
\usepackage{pgfplots}
\usepackage{pgfplotstable}

\usepackage{siunitx}
\usepackage{ifthen}
\usepackage{algorithm}
\usepackage[noend]{algpseudocode}

\usepackage{listings}
\usepackage[utf8]{inputenc}

\usepackage{booktabs}  
\usepackage{tabu}
\usepackage{array}
\usepackage{multirow}
\usepackage{diagbox}
\usepackage{tabularx}

\usepackage{balance}
\usepackage{microtype}

\usepackage[english]{babel}
\usepackage{blindtext}

\usepackage[most]{tcolorbox}

\usepackage[capitalise]{cleveref}

\def\setuppreprint{1}
\def\setupcameready{0}

\newcommand{\name}{{\sc SpliDT}\xspace}
\newcommand{\namebold}{{\sc \textbf{SpliDT}}\xspace}

\newcommand{\titleinfo}{\namebold{}: Partitioned Decision Trees for Scalable Stateful Inference at Line Rate} 

\newcommand{\authorinfo}{Murayyiam Parvez$^*$, Annus Zulfiqar$^{\diamond*}$, Roman Beltiukov$^\star$, Shir Landau Feibish$^\dagger$,\\ Walter Willinger$^\ddagger$, Arpit Gupta$^\star$, Muhammad Shahbaz$^\diamond$}

\newcommand{\institutioninfo}{
Purdue University~~~
$^\diamond$University of Michigan~~~
$^\star$UCSB~~~
$^\dagger$The Open University of Israel~~~
$^\ddagger$NIKSUN Inc.~
}
\newcommand{\submissioninfo}{\em Submission \#309 \\ {\normalsize \pageref{lastpage} Pages Body, \pageref{totalpage} Pages Total}}

\if\setupcameready1
	\def\shownames{1} 										
	\def\showpagenumbers{0}                                 
    \def\showcomments{0} 									
    \def\showacks{1} 										
\else
	\def\shownames{1}										
	\def\showpagenumbers{1}                                 
    \if\setuppreprint1										
        \def\showcomments{0}
    \else
        \def\showcomments{1}
    \fi
    \def\showacks{0}										
\fi

\newcommand{\ie}{i.e.}
\newcommand{\eg}{e.g.}

\setlist[itemize]{topsep=4pt, itemsep=4pt, parsep=1.5pt}

\definecolor{customblue}{HTML}{E6F0FA}
\definecolor{custompink}{HTML}{F7E7E7}

\if\showcomments1
    \usepackage[colorinlistoftodos,textsize=tiny]{todonotes}
\else
    \usepackage[disable]{todonotes}
\fi

\begin{document}

\title{\titleinfo}

\if\shownames1
    \author{\authorinfo \\ \institutioninfo}
\else
    \author{\submissioninfo}
\fi

\if\showcomments1
	\onecolumn
    \setcounter{page}{0}
    \listoftodos{}
    \clearpage
    \twocolumn
    \setcounter{page}{1}
\fi

\if\showpagenumbers0
	\pagestyle{empty}
\fi

\maketitle

\begin{abstract}
Machine learning (ML) is increasingly being deployed in programmable data planes (switches and SmartNICs) to enable real-time traffic analysis, security monitoring, and in-network decision-making. 
Decision trees (DTs) are particularly well-suited for these tasks due to their interpretability and compatibility with data-plane architectures, \ie, match-action tables (MATs). 
However, existing in-network DT implementations are constrained by the need to compute all input features upfront, forcing models to rely on a small, fixed set of features per flow. 
This significantly limits model accuracy and scalability under stringent hardware resource constraints.

We present \name{}, a system that rethinks DT deployment in the data plane by enabling partitioned inference over sliding windows of packets. 
\name{} introduces two key innovations: (1) it assigns distinct, variable feature sets to individual subtrees of a DT, grouped into partitions, and (2) it leverages an in-band control channel (via recirculation) to reuse data-plane resources (both stateful registers and match keys) across partitions at line rate. 
These insights allow \name{} to scale the number of stateful features a model can use without exceeding hardware limits. 
To support this architecture, \name{} incorporates a custom training and design-space exploration (DSE) framework that jointly optimizes feature allocation, tree partitioning, and DT model depth. 
Evaluation across multiple real-world datasets shows that \name{} achieves higher accuracy while supporting up to 5$\times$ more stateful features than prior approaches (\eg, NetBeacon and Leo). 
It maintains the same low time-to-detection (TTD) as these systems, while scaling to millions of flows with minimal recirculation overhead ($\le$0.05\%).
\end{abstract}

\section{Introduction}
\label{sec:intro}

\renewcommand{\thefootnote}{\fnsymbol{footnote}}
\footnotetext[1]{Both authors contributed equally to this work.}
\renewcommand{\thefootnote}{\arabic{footnote}}

Machine Learning (ML) is rapidly becoming a cornerstone of modern networking, driving increasingly sophisticated applications such as DDoS detection (LUCID~\cite{lucid}, Flowlens~\cite{flowlens}), intrusion detection~\cite{cicids17, cicids18, xNIDS}, encrypted traffic analysis~\cite{deepcorr, rosetta, ndss_traffic_analysis}, malware classification~\cite{Fu0023, malware_traffic_classification}, IoT botnet detection~\cite{horus} as well as congestion control~\cite{cubic-tcp, winstein2013tcp, yan2018pantheon, Li:2019:HHP:3341302.3342085, pcc-congestion}, and variable bitrate (VBR) video streaming~\cite{mao2017neural, puffer}.
These use cases demand real-time, high-throughput inference~\cite{taurus} to keep up with the ever-growing scale and complexity of network traffic~\cite{social_network, understanding-datacenter, cicids17, cicids18, cicvpn, ciciomt24, ciciot23}.

Programmable data planes---including {\em modern switches} (\eg, Broadcom Trident~\cite{broadcom-trident4}, Xsight X2~\cite{xsight-x2}, Intel Tofino~\cite{tofino, tofino2}, and NVIDIA Spectrum-X~\cite{nvidia-spectrumx}) and {\em emerging SmartNICs} (\eg, Intel IPU~\cite{intel-ipu-offload}, AMD Pensando DPU~\cite{amd-pensando}, and NVIDIA BlueField-3 DPU~\cite{nvidia-bluefield-dpu})---with their ability to process packets at line rate, have emerged as a powerful platform for deploying ML models directly in the network~\cite{leo, netbeacon, taurus, homunculus, iisy-hotnets, planter, mousika}.
By offloading inference tasks to network hardware, these data planes eliminate the need for control-plane intervention, enabling low-latency and high-performance decision-making~\cite{leo, netbeacon, taurus, eRSS}.

A major focus of recent work has been on mapping decision tree (DT) models to programmable data planes~\cite{iisy-hotnets, mousika, leo, netbeacon, pforest}, mainly because of these models' interpretability and natural alignment with the (reconfigurable) match-action table (MAT) architecture~\cite{tofino, tofino2, rmt, broadcom-trident4, xsight-x2, xilinx-sn-1000, xilinx-u250, broadcom-trident5, nvidia-bluefield-dpu}.
While systems such as IIsy~\cite{iisy-hotnets}, NetBeacon~\cite{netbeacon}, and Leo~\cite{leo} have demonstrated the feasibility of deploying DTs in the data plane---to be able to operate within the stringent resource constraints of these programmable switches and SmartNICs---they have mainly addressed the challenge of rule explosion and concentrated on optimizing model representations (\eg, pruning DTs~\cite{leo} and compressing MAT rules~\cite{iisy-hotnets,netbeacon}).

\begin{figure}[t]
    \centering
    \includegraphics[width=1\linewidth]{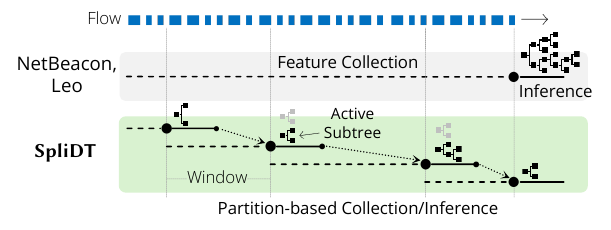}
    \vspace{-14pt}
    \caption{Comparison of in-network decision tree (DT) classification approaches. State-of-the-art methods (top) perform one-shot inference by collecting features over the entire flow duration. In contrast, \namebold{} (bottom) collects features and performs inference incrementally across partitions using windows of packets---significantly scaling the number of stateful features, while achieving higher F1 scores at line rate.}
    \label{fig:design-insight}
    \vspace{-15pt}
\end{figure}

However, despite these advancements, the critical aspect of feature collection and engineering, \ie, selecting and computing complex (stateful) input features, remains largely unexplored in this context, primarily due to the resource limitations of underlying network hardware.
Existing approaches either constrain the number of stateful features to a small, fixed set (\eg, the top-$k$ most important features, as in NetBeacon~\cite{netbeacon} and Leo~\cite{leo}) or avoid using the stateful features altogether (as in IIsy~\cite{iisy-hotnets} and Mousika~\cite{mousika}).
As we show in \S\ref{sec:background}, these strategies result in poor model performance (\eg, reduced F1 scores) and prevent deployed DT models from capturing complex, real-world traffic patterns effectively~\cite{leo, netbeacon, iisy-hotnets}.

This limited focus on feature engineering in prior work has its roots in two commonly made assumptions. 
The first assumption is that all selected features must be computed upfront before DT traversal can begin~\Cref{fig:design-insight} (top). 
Second, programmable data planes are assumed to be limited to executing DT models in a single, one-shot manner, prohibiting resource reuse across different portions of the DT. 
As a result, existing approaches view reducing the number of stateful features as the only viable solution to satisfy the given resource constraints, thus making it necessary to sacrifice model performance for scalability (\ie, supporting more concurrent flows) or vice versa~\cite{leo, netbeacon}.

In this paper, we challenge these assumptions and present \name{}, a system that enables scalable and resource-efficient deployment of DTs in the data plane by rethinking how features are computed and reused.
\name{} is built on two key insights.
First, {\em feature computation can be deferred:} DTs do not require all features to be computed upfront. 
Instead, \name{} divides the DT into partitions, where each partition---a group of consecutive layers containing one or more subtrees---computes only the features relevant to its specific subtree. 
This allows features to be computed incrementally as the DT traversal progresses, \Cref{fig:design-insight} (bottom). 
Second, {\em resources can be reused across subtrees:} by leveraging packet recirculation as an in-band control channel, \name{} reuses data-plane resources (\ie, registers and match keys) between subtrees, enabling more efficient use of constrained hardware without sacrificing line rate.

\name{} leverages these insights in an intuitive way to significantly {\em scale the total number of stateful features} that a DT can utilize. 
Instead of applying the same top-$k$ features across the entire DT, \name{} assigns each subtree---resulting from tree partitioning---its own set of relevant features, allowing feature selection to vary across subtrees. 
These subtrees are then triggered sequentially, via recirculated control packets in the data plane, reusing the stateful registers and match keys at each stage of DT traversal for the currently active subtree, \Cref{fig:design-insight} (bottom). 
We demonstrate in \S\ref{sec:evaluation} that \name{} supports five times more stateful features (\ie, the total number of unique features across all subtrees) than state-of-the-art approaches~\cite{leo, netbeacon}, all while achieving higher model accuracy and scaling to millions of concurrent flows at line rate.

In enabling these benefits, \name{} must overcome two key challenges: (1) determining how to select and compute the appropriate features at runtime for each subtree during inference, and (2) designing an effective partitioning strategy at training time that balances model accuracy and hardware resource efficiency to maximize the number of supported concurrent flows in the data plane. 

To address the first challenge, \name{} processes each flow in windows of packets, specific to each subtree of a partition.
Ideally, every subtree should have access to the entire flow during inference; however, since the data plane operates (and monitors traffic) at line rate without buffering, this is not feasible in practice~\cite{tofino, tofino2, rmt, broadcom-trident4, xsight-x2, xilinx-sn-1000, xilinx-u250}.
Instead, \name{} splits each flow into uniform windows, allowing each subtree to observe a portion of the flow during inference, \Cref{fig:design-insight} (bottom). 
Modern datacenter transport protocols (\eg, Homa~\cite{homa} and NDP~\cite{ndp}) embed flow size information in packet headers, which can be parsed in the hardware to determine window boundaries.
This information allows the data plane to halt feature collection at the designated boundary, trigger the selection of active subtrees, and transition to the next partition via recirculation.

To tackle the second challenge, \name{} employs an iterative design search methodology integrated with a custom training framework to fine-tune DT configurations, including partitioning strategies and resource allocation policies, for specific use cases (and datasets).
The goal is to optimize the trade-off between model accuracy and the number of flows that can be supported, identifying configurations that lie on the Pareto frontier. 
Leveraging Bayesian Optimization (BO), such as HyperMapper~\cite{hypermapper}, \name{} systematically explores the design space to identify the most effective hyperparameters, including the maximum number of features ($k$) per subtree, the number and size of partitions, and the overall tree depth.
Intuitively, smaller values of $k$ enable support for more flows,\footnote{In Tofino1 switch~\cite{tofino}, $k=4$ supports up to 100,000 flows, which decreases to 65,000 with $k=6$, and so on~\cite{netbeacon, leo}. SmartNICs (\eg, AMD Pensando DPU) exhibit similar behavior, with flow capacity dropping from about 64,000 ($k=4$) to 40,000 ($k=6$)~\cite{amd-pensando, nvidia-bluefield-dpu}.} while deeper subtrees generally improve model accuracy.
Similarly, increasing the number of subtrees expands the total set of unique features across the DT model, but reduces the number of packets each subtree can observe, limiting the temporal window for feature computation. 
\name{} navigates this trade-off space to derive Pareto-optimal models that balance inference accuracy and flow scalability within the constraints of the underlying hardware.

Our results (\S\ref{sec:evaluation}) demonstrate that \name{} sets a new state-of-the-art for deploying DT models in the data plane.
It consistently outperforms NetBeacon~\cite{netbeacon} and Leo~\cite{leo}, achieving a superior accuracy-to-flow count Pareto frontier across all levels of supported flows (\Cref{table:datasets}).
\name{} supports 5$\times$ more stateful features, enabling richer in-network inference, all while maintaining a low recirculation overhead---just \SI{50}{Mbps} (0.05\%) in the worst case---and matching the low time-to-detection (TTD) performance of current systems.
To facilitate reproducibility and further research in this area, we will publicly release the complete artifact, including training scripts, models, and evaluation datasets~\cite{artifact}.

\section{Background \& Motivation}
\label{sec:background}

We first review prior work on in-network classification systems that use decision tree (DT) models and highlight the need for more stateful features for DT-based inference (\S\ref{ssec:need-for-features}). 
We then revisit the anatomy of DTs to derive domain-specific insights (\S\ref{ssec:dsp-dt}) that guide us in addressing the challenges of scaling DTs on modern programmable switches (\S\ref{ssec:switch-as-tsm}).



\subsection{The Need for More Stateful Features}
\label{ssec:need-for-features}

\begin{figure}[t]
    \centering
    \includegraphics[width=1.0\linewidth]{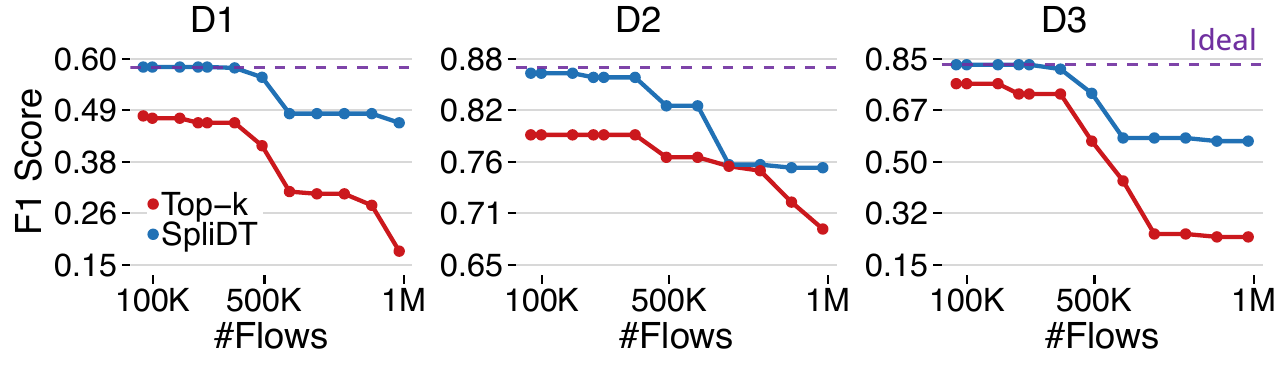}
    \vspace{-15pt}
    \caption{\namebold{} and top-$k\leq$ 7 model versus the ideal scenario with unlimited resources. \name{}, with access to all features, achieves higher F1 score than top-$k$ for the datasets, {\sf \textbf{D1--3}} (details in \S\ref{sec:evaluation}). The per-packet models peak at 0.41, 0.56, and 0.59, respectively (not shown).}
    \label{fig:motivation-top-k-pareto}
    \vspace{-15pt}
\end{figure}

As network traffic scales to multi-Tbps rates, existing approaches, such as IIsy~\cite{iisy-hotnets} and Planter~\cite{planter}, aim to support real-time inference by mapping DTs onto match-action tables (MATs) while relying solely on stateless, per-packet features. 
These methods optimize DT representation to fit within the constraints of available switch resources (\ie, MATs) but lack flow-level context, limiting their classification accuracy and adaptability~\cite{iisy-hotnets, planter, netbeacon, mousika}. 
While they efficiently handle large flow volumes in the data plane, their reliance on per-packet features significantly reduces accuracy, yielding F1 scores nearly 2$\times$ lower than models with full features access (\Cref{fig:motivation-top-k-pareto}).

More recent work, such as NetBeacon~\cite{netbeacon} and Leo~\cite{leo}, improves classification accuracy by incorporating stateful features (\ie, top-$k$), allowing DT models to leverage flow-level context, which provides richer insights than per-packet features alone. 
While this approach enhances accuracy, it places substantial pressure on the limited memory resources of programmable data planes~\cite{rmt, tofino, tofino2, broadcom-trident4, xsight-x2, xilinx-sn-1000, xilinx-u250}, ultimately limiting the number of flows that can be supported concurrently.

First, stateful features must be stored in registers, which share limited space with match-action tables (MATs) within each stage of the data-plane pipeline. 
This creates a trade-off between feature storage and model complexity. 
For example, in Tofino1~\cite{tofino}, allocating just four registers per flow exhausts an entire switch stage at 65K flows (or about 40K flows on a Pensando DPU-based SmartNIC~\cite{amd-pensando}), preventing that stage from being used for model execution. 
Increasing the number of registers or supporting more flows further reduces available MAT stages, limiting DT depth and restricting feature selection. 
Second, adding more stateful features increases match key sizes, which inflates the size of table entries and exacerbates TCAM memory usage---making it harder to map DTs efficiently onto MATs in the data plane~\cite{leo, netbeacon, tofino}.

As a result, prior work has been limited to a maximum of 200K flow rules and a handful of stateful features (top-$k \leq 6$)---bounds imposed by hardware memory constraints and the need to balance register usage with model depth---yielding only moderate accuracy gains~\cite{leo, netbeacon}. 
Beyond this threshold, performance degrades: DT depth becomes restricted, lowering classification accuracy, while larger match key sizes increase TCAM overhead, reducing scalability (\S\ref{sec:evaluation}). 
These trade-offs highlight the core challenge of integrating stateful features into DT-based inference without sacrificing scalability on resource-constrained programmable data planes.

\vspace{5pt}\noindent
\textbf{\em Observation:}
The constraints of prior DT-based systems are often perceived as intrinsic to programmable data-plane architectures, but are they truly fundamental? 
We argue that these limitations do not stem from hardware constraints but from ingrained assumptions about how DTs should be processed. 
Conventional approaches assume that all stateful features must be collected before inference begins, necessitating upfront register allocation. 
Furthermore, these approaches treat DT execution as a single-pass, feed-forward process, confining computation to the spatially available pipeline resources.

This paper challenges the prevailing belief that feature richness and scalability must always be in conflict. 
We demonstrate that by decoupling stateful feature selection from DT execution, both can scale independently. 
Unlike prior work (\eg, NetBeacon~\cite{netbeacon} and Leo~\cite{leo}), which sacrifices feature expressiveness for flow scalability, \name{} dynamically selects and reuses stateful features across inference steps, efficiently managing available hardware resources without restricting model complexity. 
Compared to traditional top-$k$ systems, \name{} achieves a significantly improved Pareto frontier, simultaneously enhancing F1 scores and flow rule capacity (\Cref{fig:motivation-top-k-pareto}). 
These results challenge the notion that hardware-imposed constraints inherently limit DT scalability, showing instead that the primary bottleneck arises from rigid execution models that preallocate features and enforce single-pass processing. 

\subsection{Domain-Specific Properties of DTs}
\label{ssec:dsp-dt}

\begin{figure}[t]
    \centering
    \includegraphics[width=0.92\linewidth]{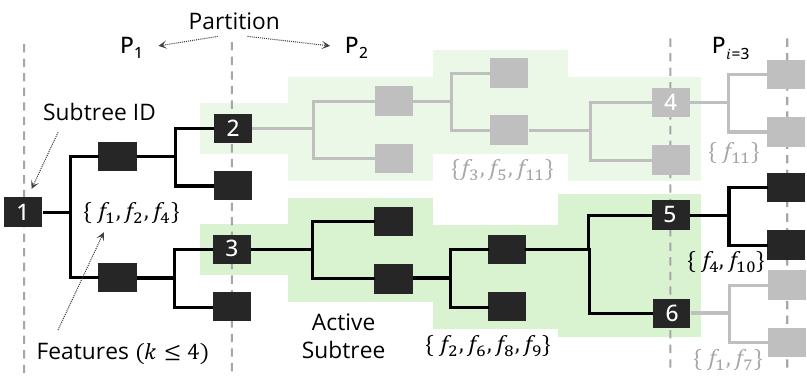}
   \vspace{-5pt}
    \caption{Domain-specific properties of DTs: Partitions ($P_i$) comprising multiple subtrees, each with its subset of features ($k$). During traversal, active subtrees are selected within each partition.}
    \label{fig:partitioned-dt}
    \vspace{-18pt}
\end{figure}

DT inference begins at the root node, where a decision is made based on selected features to determine the next node to visit. 
This process continues at each level, using different features at each step until a leaf node is reached. 
Instead of processing the tree level by level, we can group consecutive levels into {\em partitions} (\Cref{fig:partitioned-dt}) and focus on the subtrees in each partition. 
With this approach, inference progresses one partition at a time, where the decisions resulting from traversing the {\em active} subtree in one partition determine which subtree to traverse in the next partition.
This allows for more efficient traversal, selecting only the next subtree based on relevant features and their conditions rather than evaluating entire levels sequentially.

This subtree-by-subtree execution enables features to be collected incrementally and on-demand for the active subtree in each partition (\Cref{fig:design-insight}, bottom). 
Unlike traditional approaches that require gathering all features upfront, this method loads only the features needed for the current subtree. 
With just $k$ available feature slots, we can dynamically allocate and process only the relevant features at each step, avoiding the restrictive top-$k$ selection enforced by existing systems~\cite{leo, netbeacon}. 
This approach maximizes feature utilization without discarding valuable contextual information across the entire DT.

However, two key conditions must hold for this approach to be effective: (a) {\em feature density across subtrees}: each subtree within a partition must use at most $k$ features out of the total $N$ available features. 
Unlike traditional top-$k$ approaches that select a fixed set of $k$ features for the entire tree, here $k$ represents the number of feature slots available at any given step (subtree), which can be dynamically reassigned to different features as the inference progresses. 
If even a single subtree requires more than $k$ features, the benefits of incremental feature computation are lost, as more than $k$ features would need to be allocated upfront; (b) {\em incremental computation architecture}: the system must support subtree-by-subtree traversal, dynamically collecting and computing features while reusing the same $k$ feature slots at each step. 
This ensures efficient execution without requiring all features to be preloaded simultaneously, overcoming the limitations of prior top-$k$ selection methods that discard all but the most globally important features across the entire DT.

To examine the feasibility of the first condition, we studied feature usage across subtrees in multiple datasets (\Cref{table:motivation-feature-dispersion-and-resubmission}). 
In the considered datasets ({\sf D1--3}),\footnote{\label{footnote1}This trend extends beyond {\sf D1--3}; similar feature sparsity holds for {\sf D4--7} (see~\S\ref{sec:evaluation}) and is commonly observed in real-world DT classification tasks~\cite{malware_traffic_classification, netfound}.} we found that at most only 10\% of features were required in any given subtree.
For instance, in dataset {\sf D1}, where $N=41$, subtrees required on average only 3.73 features. 
Note that this 11$\times$ reduction in storage requirement makes it possible to execute the entire DT using only $k$ registers (\eg, 4 for {\sf D1}) and avoids the need to impose a strict limit of $k$ on the overall features.

For the second condition, we demonstrate in the next section (\S\ref{ssec:switch-as-tsm}) how modern programmable architectures, with support for packet recirculation~\cite{tofino, tofino2, broadcom-trident4, xsight-x2, xilinx-sn-1000, xilinx-u250}, can be reimagined as time-shared machines, enabling efficient reuse of resources (\eg, registers and match keys) across partitions within the data plane.

\begin{table}[t]
\footnotesize
\begin{tabularx}{\linewidth}{X|r|r|r|r}

\toprule

\multirow{2}{*}{\bf Data}
& \multicolumn{2}{c|}{\bf Feature Density (\%)}
& \multicolumn{2}{c}{\bf Recirc. Bandwidth (Mbps)} \\

& {\bf / Partition}
& {\bf / Subtree}
& \multicolumn{1}{c}{\bf WS}
& \multicolumn{1}{c}{\bf HD} \\

\toprule

{\sf D1}
& 47.15\,$\pm$\,38.44
& 6.15\,$\pm$\,2.95
& 2.93\,$\pm$\,2.44
& 5.99\,$\pm$\,3.51 \\

{\sf D2}
& 53.49\,$\pm$\,44.19
& 7.28\,$\pm$\,2.72
& 6.01\,$\pm$\,4.01
& 12.32\,$\pm$\,5.76 \\

{\sf D3}
& 53.95\,$\pm$\,43.42
& 6.08\,$\pm$\,3.37
& 3.58\,$\pm$\,3.21
& 7.33\,$\pm$\,4.62 \\

\bottomrule
\end{tabularx}
\vspace{-4pt}
\caption{Feature density (\%) across partitions and subtrees in a DT, and max. recirculation bandwidth (Mbps) when processing datasets ({\sf \textbf{D1--3}}) for two datacenter environments, Webserver (WS) and Hadoop (HD), \S\ref{sec:evaluation}.}
\vspace{-16pt}
\label{table:motivation-feature-dispersion-and-resubmission}
\end{table}

\subsection{Switch as a Time-Shared Machine}
\label{ssec:switch-as-tsm}

Programmable data planes are traditionally viewed as spatial architectures with fixed resource constraints, where exceeding available resources leads to failures during program compilation~\cite{p4-paper, rmt, tofino}.
However, we argue that because this perspective focuses mainly on the static aspects of both programmable data planes and spatial architectures, it is overly restrictive and in need of being revisited.

For example, modern data planes (switches~\cite{tofino, tofino2} and SmartNICs~\cite{xilinx-sn-1000, xilinx-u250}) support packet recirculation, with bandwidths reaching \SI{100}{Gbps} (\eg, Tofino~\cite{tofino}, X2~\cite{xsight-x2}, and Trident~\cite{broadcom-trident4}), without impacting line rate.
This capability adds an important dynamic element to the traditional view. 
In effect, it enables temporal execution, allowing different stages of a program (\eg, in P4~\cite{p4-paper} or NPL~\cite{npl}) to activate across multiple recirculations in a time-shared manner. 
By leveraging this mechanism, the state can be distributed over time, facilitating the reuse of limited resources (\eg, registers and match keys in MATs).\footnote{Recirculation in \name{} is used as a fast, in-band control channel---not for forwarding data traffic---avoiding the overhead of the software control plane.}

\begin{figure*}[t]
    \centering
    \includegraphics[width=\linewidth]{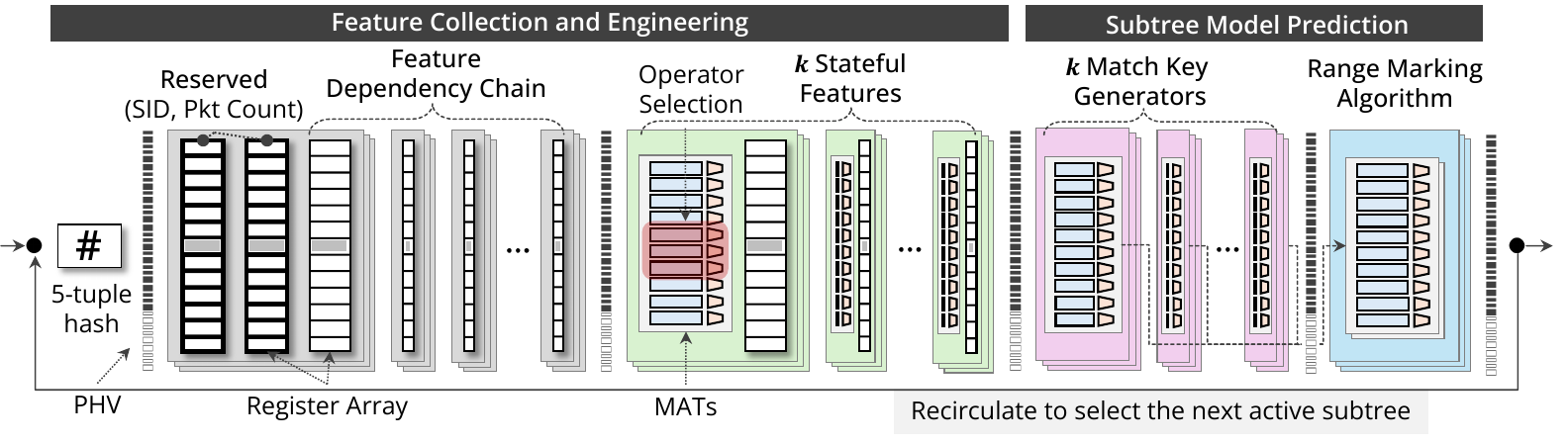}
    \vspace{-14pt}
    \caption{\namebold{}'s partitioned inference architecture, processing flow windows in two phases: (1) Feature Collection and Engineering (left) and (2) Subtree Model Prediction (right)---leveraging resource reuse (\ie, registers and match keys) via recirculation, for efficient execution within each DT partition.}
    \label{fig:part-inf-arch}
    \vspace{-14pt}
\end{figure*}

By carefully restructuring DTs (\eg, expressed in P4~\cite{p4-paper} or NPL~\cite{npl}), we can exploit this dynamic capability to scale DT-based inference beyond the physical limits of available resources---akin to how CPUs abstract resource constraints by reusing registers over time. 
\Cref{table:motivation-feature-dispersion-and-resubmission} shows that for the evaluated datasets ({\sf D1--3}),\footnotemark[\getrefnumber{footnote1}] the maximum recirculation path usage is within \SI{20}{Mbps} for the two datacenter environments E1--2, significantly lower than the total available bandwidth of \SI{100}{Gbps}.

In the next section, we show how \name{} leverages the domain-specific properties of DTs and reuse of switch resources (via recirculation) to optimize the Pareto frontier (F1 score vs. number of flow rules) while supporting all stateful features at line rate.

\section{Design of \namebold{}}
\label{sec:design}

We now present the \name{} design; its {\em partitioned inference architecture} for efficient DT execution in the data plane (\S\ref{ssec:partitioned-inference-architecture}) and the {\em custom training framework}, for effective model configuration through design search (\S\ref{ssec:custom-training}).
We then bring these components together to show how \name{} enables scalable and resource-efficient inference in practice (\S\ref{ssec:putting-together}).

\subsection{Partitioned Inference Architecture}
\label{ssec:partitioned-inference-architecture}

As illustrated in \Cref{fig:part-inf-arch}, \name{}'s partitioned inference architecture operates in two phases: (1) {\em Feature Collection and Engineering} (\S\ref{sssec:feature-collection}) and (2) {\em Subtree Model Prediction} (\S\ref{sssec:st-model-prediction}).
For each flow, it iteratively processes windows of packets using the active subtree within each DT partition, leveraging {\em resource reuse} (\ie, registers and match keys) via recirculation (\S\ref{sssec:resource-reuse}).

\subsubsection{Feature Collection and Engineering.}
\label{sssec:feature-collection}

In \name{}, we maintain three distinct register-array sets to manage stateful feature collection and subtree selection for prediction, as shown in \Cref{fig:part-inf-arch}. 
These registers include: (1) reserved state registers for tracking metadata such as the subtree ID (SID) and per flow packet counters, (2) registers for computing intermediate and dependent states (\eg, timestamps for inter-arrival time (IAT) calculations), and (3) $k$ registers for storing stateful features specific to the active subtree within the current DT partition.

\begin{tcolorbox}[
    colback=blue!5!white,
    colframe=black,
    title={},
    fonttitle=\bfseries,
    fontupper=\footnotesize, 
    sharp corners,
    boxrule=0.4mm,
    coltitle=black,
    enhanced jigsaw,
    drop shadow={black!50!white}
]
\textbf{INFO:} \textit{Reserved and dependency chain registers can significantly limit the number of features per subtree ($k$) as they must scale alongside the $k$ features to support the same number of flows. 
This contention for register space creates a tradeoff that must be carefully managed to balance feature capacity and flow scalability in \name{}, \S\ref{ssec:custom-training}.}
\end{tcolorbox}

Upon packet arrival, \name{} hashes its 5-tuple using CRC32~\cite{crc32} to determine the register index corresponding to the flow. 
First, it retrieves the subtree ID (SID) from the reserved register array and updates the packet count in the second register array. 
Next, depending on the use case (and dataset), some DT models require intermediate values to compute stateful features before prediction. 
For instance, IAT computation requires storing the previous packet's timestamp to calculate the inter-packet gap.

To support such computations, \name{} implements a dependency chain, a sequence of register arrays distributed across multiple pipeline stages to enable hierarchical computation. 
Since programmable data planes~\cite{tofino, tofino2, rmt, broadcom-trident4,broadcom-trident5, nvidia-bluefield-dpu} cannot process dependent data within a single stage, computation must be spread across multiple stages. 
In our evaluations, the deepest observed dependency chain was 3 stages---a depth well within the capabilities of modern data-plane architectures.

\vspace{5pt}\noindent
\textbf{\em Operator Selection.}
Since each subtree may require a different operation to compute its stateful features, \name{} dynamically updates the operation applied to each feature in every flow window. 
To achieve this, \name{} utilizes match-action tables (MATs) for each stateful feature, acting as selectors to apply the appropriate operation on demand (\Cref{fig:part-inf-arch}).

At compile time (\S\ref{ssec:design-search}), \name{} populates these tables with rules that define which operator to apply for each subtree. 
The MATs match on the subtree ID (SID) and select the corresponding action to perform the necessary computation. 
Modern data-plane architectures support the parallel execution of multiple MATs within a stage. 
For instance, Tofino1 supports up to 16 MATs with 750 entries each, which is well within the requirements of our design---\name{} utilizes only six MATs to support $k=6$ stateful features in our evaluations (\S\ref{sec:evaluation}), with each table containing at most 200 entries.

Lastly, to select the appropriate features to populate in the feature registers, prevent continuous feature updates on every arriving packet, and identify window boundaries, \name{} incorporates additional match fields in the MATs. 
For instance, to update a stateful feature only on SYN packets (such as SYN packet count), the MATs can include TCP flags as a match condition, ensuring the feature update is triggered only when a SYN is received.

\subsubsection{Subtree Model Prediction.}
\label{sssec:st-model-prediction}
~The prediction phase (\Cref{fig:part-inf-arch}) executes the decision tree (DT) model using the Range Marking Algorithm~\cite{netbeacon}, which maps the partitioned DT into a sequence of MATs.
The first group of MATs, the $k$ match key generators, constructs the match keys required for the DT model.
As with the operator selection tables, \name{} maintains a separate MAT for each feature, using precomputed rules to determine which stateful feature (register) serves as the key.
Current feature values---accumulated over the most recent flow window---are stored in dedicated metadata headers.
These metadata values are then used as match keys in the generator tables.
Each table's action produces a range mark, a unique bit string written to a corresponding metadata field.
These per-feature range marks, combined with the subtree ID (SID), form the match keys for a final MAT that implements the Range Marking Algorithm, encoding the DT model rules and performing classification.

If the current subtree is in an intermediate partition, classification yields the next SID; if it is in the final partition or an early-exit node, it outputs the flow's final class label.
In the latter case, the label is sent to the controller as a digest.
Otherwise, the next SID is recirculated via the resubmission channel (\S\ref{sssec:resource-reuse}), updating the SID register and resetting the dependency chain and $k$ stateful feature registers in preparation for the next flow window.

\begin{figure}[t]
    \centering
    \includegraphics[width=0.90\linewidth]{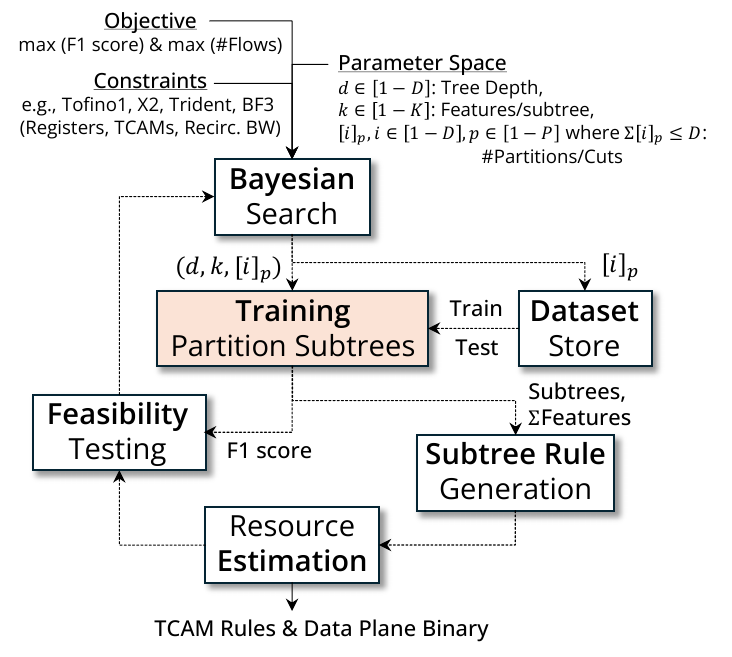}
    \vspace{-4pt}
    \caption{Workflow of \namebold{}'s Model Design Search.}
    \label{fig:design-search}
    \vspace{-18pt}
\end{figure}

\subsubsection{Resource Reuse via In-Band Control Channel (Recirculation).}
\label{sssec:resource-reuse}
At the end of inference for each subtree, the next subtree ID (SID) is propagated to the stateful registers in the feature collection and engineering stages via resubmission, which serves as an in-band control channel~\cite{tofino}.
A single packet---triggered after processing the corresponding flow window---is resubmitted with the updated SID encoded in a metadata header field.
This resubmission imposes minimal bandwidth overhead (see~\Cref{table:motivation-feature-dispersion-and-resubmission}), as only one control packet is required per flow window.
This mechanism allows \name{} to perform partitioned DT inference incrementally, reusing the same pipeline resources across subtrees without introducing resource conflicts or execution overhead.

\subsection{Custom Search/Training Framework}
\label{ssec:custom-training}

We now describe \name{}'s design search framework for generating optimal DTs (\S\ref{ssec:design-search}) and partitioning them into subtrees using a custom training algorithm (\S\ref{ssec:partitioned-training}).

\subsubsection{\namebold{} Design Search.}
\label{ssec:design-search}
~\Cref{fig:design-search} illustrates the overall workflow of the \name{} framework.
The goal is to generate a Pareto frontier of partitioned DT configurations.
Given a set of optimization objectives, input parameter ranges, and hardware/performance constraints, a Bayesian Optimization (BO) search phase begins, iteratively proposing model configurations for evaluation.
For each configuration, we query a corresponding window-based training/test dataset based on the suggested number of partitions,\footnote{These datasets are preprocessed offline and can be efficiently stored and queried using commercial databases, \eg, PostgreSQL~\cite{postgresql} or MongoDB~\cite{mongodb}.} and use \name{}'s custom partitioned DT training algorithm (\S\ref{ssec:partitioned-training}) to train the model and evaluate its F1 score.
We then generate the corresponding TCAM entries, compute hardware resource usage, determine flow scalability, and assess whether the model can be deployed at line rate on the target switch.
The results---F1 score, supported flow count, and feasibility metrics---are fed back into the BO loop to guide the next iteration; this process continues for a predefined number of iterations.

In the following sections, we detail each stage of this workflow, beginning with the inputs to the BO search.

\vspace{5pt}\noindent
\textbf{\em Parameter Space, Objectives, and Constraints.}
To initiate the design search, the user specifies three key components that define the configuration space and guide the optimization:

\begin{itemize}[leftmargin=*]
\item {\em Model Hyperparameters:}
These define the structure and complexity of the partitioned DT. 
The search space includes: the maximum tree depth ($D$), the number of features per subtree ($k$), and a list of partition sizes $[i_1, i_2, \dots, i_p]$, where $p$ is the number of partitions and the sum of the partition sizes equals the total tree depth, \ie, $D = \sum[i_1, i_2, \dots, i_p]$.
\item {\em Hardware and Performance Constraints:}
These reflect the capabilities of the target platform (switch or SmartNIC), including available TCAM blocks, register space, pipeline stages, and recirculation bandwidth. 
They ensure that candidate models are not only accurate but also deployable within the hardware's resource envelope, at line rate.
\item {\em Optimization Objectives:}
The design search seeks to jointly maximize model accuracy (\eg, F1 score) and flow scalability (\eg, number of concurrent flows), yielding a Pareto frontier of configurations that effectively trade off between these two metrics.
\end{itemize}

Together, these inputs, along with the dataset specification, are fed into the Bayesian Optimization (BO) loop that drives the search for feasible and high-performing DT configurations for \name{}.

\vspace{5pt}\noindent
\textbf{\em Bayesian Search.}
Given the input search space and objectives, the framework performs a design-space exploration to generate a Pareto frontier of \name{} DT configurations for a target dataset.
Bayesian Optimization (BO) is a black-box optimization method~\cite{bayesian-optimization} designed for optimizing expensive-to-evaluate functions~\cite{hypermapper, smac3, bohb}.
It constructs a probabilistic surrogate model, typically a Gaussian Process (GP) or Random Forest, to approximate the objective function, and employs an acquisition function to guide the selection of promising candidates.
By balancing exploration (sampling new regions) and exploitation (refining high-performing regions), BO efficiently searches high-dimensional spaces with minimal evaluations---ideal for costly tasks such as hyperparameter tuning or model training~\cite{hyper_tuning}.

At each step, the BO search proposes several DT configurations to evaluate in parallel.
Each configuration is used to train a partitioned DT using \name{}'s custom training algorithm (\S\ref{ssec:partitioned-training}), which is then evaluated on a test dataset.
Subsequent stages assess the model's hardware resource usage, deployment feasibility, and supported flow count; these metrics are fed back to the BO loop to inform the next iteration.
This process continues for a predetermined number of iterations.

\vspace{5pt}\noindent
\textbf{\em Subtree Rule Generation.}
Given a trained partitioned DT, we generate the corresponding TCAM rules to represent the model in the data plane.
We adopt the Range Marking algorithm~\cite{netbeacon}, which efficiently encodes decision tree rules by mapping feature value ranges to compact ternary bit strings.
It segments each feature's domain into non-overlapping ranges and assigns a unique range mark (bit string) to each, ensuring that merged ranges retain distinct and unambiguous encodings.
For each feature, its thresholds from the trained partitioned DT are translated into ternary matches, with the associated range mark output actions (\ie, using P4).
These TCAM entries are installed in feature tables, producing range marks as match keys for the subsequent model table (\Cref{fig:part-inf-arch}).

A second set of TCAM rules is then generated to encode the model logic: these rules match on feature range marks and return either the next subtree ID (for intermediate partitions) or the final prediction class (at the last partition).
This encoding maps each DT leaf to a single TCAM rule, effectively avoiding rule explosion.

Both feature and model TCAM entries are generated for each subtree in the partitioned DT and installed into the corresponding tables in the data-plane pipeline (\Cref{fig:part-inf-arch}).
Each rule also includes an exact match on the subtree ID (SID) to ensure the correct subtree is selected for each partition.

\vspace{5pt}\noindent
\textbf{\em Resource Estimation and Feasibility Testing.}
To evaluate each candidate configuration, we estimate the number of TCAM blocks and pipeline stages required for the feature and model tables using a target-specific analytical model (\eg, BF-SDE~\cite{intel-p4studio}, P4Insight~\cite{intel-p4-insight}, NetASM~\cite{netasm}, and Nvidia DOCA P4 Developer Toolkit~\cite{doca}).
We then determine the number of remaining stages available for register allocation and bookkeeping logic, which directly impacts the number of flows the model can support concurrently.
Recirculation overhead, in terms of in-band control traffic, is estimated using: (1) the number of partitions, which dictates the number of recirculated packets per flow; (2) flow-size distribution observed in real-world datacenter workloads (\S\ref{sec:evaluation}); and (3) the number of active flows concurrently issuing recirculations.
A design is deemed feasible if it fits within the target's TCAM, register, and MAT stage budgets while keeping recirculation traffic within available bandwidth limits.

The feasibility outcome (yes/no), along with the model's F1 score and supported flow count, is fed back into the BO loop to guide subsequent iterations.
At each step, the BO search proposes new parameters (and configurations), gradually converging on models that jointly maximize accuracy and flow scalability, while satisfying the target resource constraints.

\begin{figure}[t]
\vspace{-8pt}
\centering
\setlength{\abovecaptionskip}{-5pt}
\setlength{\belowcaptionskip}{0pt}

\begin{algorithm}[H]
   \footnotesize 
    \caption{\name{} partitioned DT training algorithm.}
    \label{fig:partitioned-training}
    \begin{algorithmic}[1]
    \Procedure{TrainPartDT}{$dataset, depths, partition, k$}
    \State \textcolor{gray}{/* k is the features per subtree */}
    
    \If{$partition \geq$ \textbf{len}($depths$)}
        \State \textbf{return}
    \EndIf
    \State $depth \gets depths[partition]$

    \State \textcolor{gray}{/* train one subtree at this partition */}
    \State $tree \gets$ TrainSubTree($dataset[partition], depth, k$)

    \State \textcolor{gray}{/* get subset of data samples for each leaf node */}
    \State $leaf\_subsets \gets$ PartitionSamplesByLeaves($tree, dataset$)

    \State \textcolor{gray}{/* train subtrees for the next partition */}
    \ForAll{$(leaf, subset) \in leaf\_subsets$}
        \State \textcolor{gray}{/* train only if not an exit node */}
        \If{\textbf{len}($subset$) $>$ 0}
            \State TrainRecursiveTree($subset, depths, partition + 1$)
        \EndIf
    \EndFor
    \EndProcedure
    \end{algorithmic}
\end{algorithm}
\vspace{-25pt}
\end{figure}

\subsubsection{\namebold{} Custom Training.}
\label{ssec:partitioned-training}
\Cref{fig:partitioned-training} outlines the algorithm used to train \name{}'s partitioned decision trees.
Given the overall tree depth, partition sizes, and the number of features per subtree, training begins by learning a single subtree for the first partition using all samples from the corresponding initial window.
For each leaf node in this subtree, we identify the subset of training samples that reach that node and use only those samples---along with their associated window for the next partition---to train the corresponding subtree in the following partition.
Leaf nodes that do not reach the maximum depth in a partition do not generate further subtrees.
This recursive approach allows each subtree to specialize based on the subset of flow packets it receives, enabling window-based inference matching the data distribution observed during training.

\subsection{Putting It All Together}
\label{ssec:putting-together}

\begin{table*}[t]
\footnotesize
\centering
\begin{tabularx}{\textwidth}{lXrr}

\toprule

{\bf Dataset} & {\bf Description} & {\bf Classes} \\
\toprule

\textsf{D1}: \textsf{CIC-IoMT2024} & A cybersecurity dataset~\cite{ciciomt24} with Internet of Medical Things (IoMT) traffic for intrusion detection in healthcare. & 19 \\
\textsf{D2}: \textsf{CIC-IoT2023-a} & A simplified version of the CIC-IoT-2023 dataset~\cite{ciciot23}, categorized into four primary classes of IoT traffic. & 4 \\
\textsf{D3}: \textsf{ISCX-VPN2016} & A dataset containing VPN and non-VPN traffic~\cite{cicvpn} for evaluating VPN detection and privacy-related analyses. & 13 \\
\textsf{D4}: \textsf{CampusTraffic} & UCSB campus dataset~\cite{netfound} containing various application types, including web, cloud, social, and, streaming traffic. & 11 \\
\textsf{D5}: \textsf{CIC-IoT2023-b} & A comprehensive IoT dataset~\cite{ciciot23} containing multi-class network traffic data for evaluating IoT security threats. & 32 \\
\textsf{D6}: \textsf{CIC-IDS2017} & A network intrusion detection dataset~\cite{cicids17} for various attack scenarios, including DoS, DDoS, and brute force. & 10 \\
\textsf{D7}: \textsf{CIC-IDS2018} & An anomaly detection dataset~\cite{cicids18} capturing network traffic for diverse attacks and benign activities. & 10 \\
\bottomrule
\end{tabularx}
\vspace{-8pt}
\caption{Real-world network traffic datasets used for evaluating \namebold{} across diverse security scenarios.}
\label{table:datasets}
\vspace{-14pt}
\end{table*}

To demonstrate how \name{}'s design search and partitioned inference architecture operate in practice, we walk through a complete end-to-end example.

The process begins with the user providing a labeled dataset, model hyperparameters, resource constraints, and optimization objectives (\S\ref{ssec:design-search}).
\name{} then invokes Bayesian Optimization (BO) to explore the configuration space and identify a Pareto-optimal model that jointly maximizes F1 score and flow scalability.
For one such data point (\Cref{fig:partitioned-dt}), the selected model had a tree depth of $D=6$, with 4 features per subtree and partition sizes of $[2, 3, 1]$ across 3 partitions, supporting up to 1~Million concurrent flows.

Given this configuration (\Cref{fig:partitioned-dt}), \name{} uses its custom training algorithm (\S\ref{ssec:partitioned-training}) to train 6 subtrees: subtree 1 in the first partition, subtrees 2--3 in the second, and subtrees 4--6 in the third. 
Some subtrees are skipped due to early exits.
\name{} then generates TCAM rules for each subtree using the Range Marking Algorithm~\cite{netbeacon}, and compiles target-specific code (\eg, P4) for feature extraction and inference logic.
These rules are installed into the data plane's MATs: feature tables encode stateful features into range marks, while model tables use them to either classify flows or select the next subtree (\Cref{fig:part-inf-arch}).
All subtrees are present on the switch, but only one is active per flow at a time, requiring just 4 feature registers per flow.

Referring to the example in~\Cref{fig:partitioned-dt}, at runtime, a new flow begins with subtree ID, SID~=~1, in partition $P_1$.
The switch collects the features required by this subtree (\eg, $f_1$, $f_2$, $f_4$) over the first window of packets.
Once the window concludes, the subtree predicts SID~=~3 in partition $P_2$, prompting a single recirculation that updates the flow's SID and clears its feature and dependency-chain registers.
The second window is processed using the features needed by subtree 3 (\eg, $f_2$, $f_6$, $f_8$, $f_9$).
Following this, the model selects subtree 5 in partition $P_3$, and another recirculation updates the SID~=~5.
The third window is then processed with the features required by subtree 5 (\eg, $f_4$, $f_{10}$), and the final prediction is emitted. 
No further recirculations are needed.

This pipeline executes independently per flow, with register state indexed via 5-tuple hashing to avoid conflicts.
In this example, \name{} supports 1~Million concurrent flows, each maintaining only 4 stateful feature registers and a single SID register---demonstrating its scalability and efficiency under practical resource constraints.

\section{Implementation}
\label{sec:implementation}

\begin{figure*}[t]
    \centering
    \includegraphics[width=1.0\linewidth]{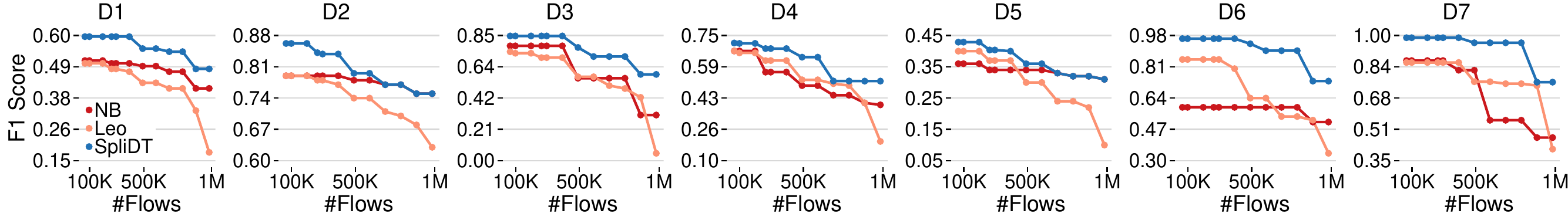}
    \vspace{-16pt}
    \caption{Pareto frontier of \namebold{} vs. baselines, indicating the best F1 score for a given \#flows in the data plane.}
    \label{fig:pareto}
    \vspace{-2pt}
\end{figure*}

\begin{table*}[t]
\footnotesize
\centering

\resizebox{0.98\textwidth}{!}{ 

\begin{tabular}{cc !{\vrule width 0.6pt} rrr !{\vrule width 0.6pt} rrr !{\vrule width 0.1pt} rrr !{\vrule width 0.6pt} rrr !{\vrule width 0.1pt} rrr} 

\toprule

\multirow{2}{*}{\bf Data} & 
\multirow{2}{*}{\bf \#Flows} & 
\multicolumn{3}{c !{\vrule width 0.6pt} }{\bf F1 Score} &  
\multicolumn{3}{c !{\vrule width 0.1pt} }{\bf Depth~/~\#Partitions} & 
\multicolumn{3}{c !{\vrule width 0.6pt} }{\bf \#Features} & 
\multicolumn{3}{c !{\vrule width 0.1pt} }{\bf \#TCAM Entries} & 
\multicolumn{3}{c}{\bf Register Size (bits)} \\ 

& & 

\multicolumn{1}{c}{NB} & 
\multicolumn{1}{c}{Leo} & 
\multicolumn{1}{c !{\vrule width 0.6pt} }{\bf \cellcolor{customblue}\name{}} &

\multicolumn{1}{c}{NB} & 
\multicolumn{1}{c}{Leo} & 
\multicolumn{1}{c !{\vrule width 0.1pt} }{\bf \cellcolor{customblue}\name{}} &

\multicolumn{1}{c}{NB} & 
\multicolumn{1}{c}{Leo} & 
\multicolumn{1}{c !{\vrule width 0.6pt} }{\bf \cellcolor{customblue}\name{}} &

\multicolumn{1}{c}{NB} & 
\multicolumn{1}{c}{Leo} & 
\multicolumn{1}{c !{\vrule width 0.1pt} }{\bf \cellcolor{customblue}\name{}} &

\multicolumn{1}{c}{NB} & 
\multicolumn{1}{c}{Leo} & 
\multicolumn{1}{c}{\bf \cellcolor{customblue}\name{}} \\

\toprule

\multirow{3}{*}{\sf D1} & 100K & 0.51 & 0.50 & \cellcolor{customblue}0.60 & 13 & 11 & \cellcolor{customblue}{24~/~5} & 6 & 6 & \cellcolor{customblue}20 & 6,596 & 16,384 & \cellcolor{customblue}4,583 & 192 & 192 & \cellcolor{customblue}160 \\
			& 500K & 0.49 & 0.43 & \cellcolor{customblue}0.55 & 13 & 6 & \cellcolor{customblue}{21~/~5} & 4 & 4 & \cellcolor{customblue}21 & 6,154 & 2,048 & \cellcolor{customblue}4,509 & 128 & 128 & \cellcolor{customblue}96 \\
			& 1M & 0.41 & 0.18 & \cellcolor{customblue}0.48 & 12 & 3 & \cellcolor{customblue}{13~/~1} & 2 & 2 & \cellcolor{customblue}2 & 1,534 & 2,048 & \cellcolor{customblue}8,238 & 64 & 64 & \cellcolor{customblue}64 \\
\midrule
\multirow{3}{*}{\sf D2} & 100K & 0.79 & 0.79 & \cellcolor{customblue}0.86 & 12 & 11 & \cellcolor{customblue}{45~/~3} & 6 & 6 & \cellcolor{customblue}30 & 8,479 & 16,384 & \cellcolor{customblue}17,083 & 192 & 192 & \cellcolor{customblue}160 \\
			& 500K & 0.78 & 0.74 & \cellcolor{customblue}0.80 & 12 & 6 & \cellcolor{customblue}{27~/~3} & 2 & 4 & \cellcolor{customblue}30 & 11,996 & 2,048 & \cellcolor{customblue}12,002 & 64 & 128 & \cellcolor{customblue}64 \\
			& 1M & 0.75 & 0.63 & \cellcolor{customblue}0.75 & 18 & 3 & \cellcolor{customblue}{12~/~1} & 1 & 2 & \cellcolor{customblue}2 & 2,540 & 2,048 & \cellcolor{customblue}465 & 32 & 64 & \cellcolor{customblue}64 \\
\midrule
\multirow{3}{*}{\sf D3} & 100K & 0.78 & 0.73 & \cellcolor{customblue}0.85 & 13 & 11 & \cellcolor{customblue}{21~/~2} & 6 & 5 & \cellcolor{customblue}23 & 5,056 & 16,384 & \cellcolor{customblue}2,562 & 192 & 160 & \cellcolor{customblue}160 \\
			& 500K & 0.56 & 0.57 & \cellcolor{customblue}0.77 & 12 & 6 & \cellcolor{customblue}{16~/~5} & 4 & 4 & \cellcolor{customblue}28 & 1,566 & 2,048 & \cellcolor{customblue}2,931 & 128 & 128 & \cellcolor{customblue}64 \\
			& 1M & 0.31 & 0.05 & \cellcolor{customblue}0.59 & 4 & 3 & \cellcolor{customblue}{18~/~4} & 2 & 1 & \cellcolor{customblue}15 & 170 & 2,048 & \cellcolor{customblue}928 & 64 & 32 & \cellcolor{customblue}32 \\
\midrule
\multirow{3}{*}{\sf D4} & 100K & 0.67 & 0.66 & \cellcolor{customblue}0.71 & 13 & 10 & \cellcolor{customblue}{28~/~2} & 6 & 7 & \cellcolor{customblue}27 & 11,361 & 8,192 & \cellcolor{customblue}14,507 & 192 & 224 & \cellcolor{customblue}160 \\
			& 500K & 0.49 & 0.52 & \cellcolor{customblue}0.64 & 10 & 11 & \cellcolor{customblue}{20~/~2} & 4 & 2 & \cellcolor{customblue}26 & 4,355 & 16,384 & \cellcolor{customblue}12,934 & 128 & 64 & \cellcolor{customblue}64 \\
			& 1M & 0.39 & 0.20 & \cellcolor{customblue}0.51 & 9 & 3 & \cellcolor{customblue}{27~/~2} & 2 & 1 & \cellcolor{customblue}1 & 4,221 & 2,048 & \cellcolor{customblue}393 & 64 & 32 & \cellcolor{customblue}32 \\
\midrule
\multirow{3}{*}{\sf D5} & 100K & 0.36 & 0.40 & \cellcolor{customblue}0.43 & 10 & 10 & \cellcolor{customblue}{49~/~5} & 6 & 7 & \cellcolor{customblue}20 & 7,755 & 8,192 & \cellcolor{customblue}27,302 & 192 & 224 & \cellcolor{customblue}96 \\
			& 500K & 0.34 & 0.30 & \cellcolor{customblue}0.36 & 12 & 6 & \cellcolor{customblue}{33~/~4} & 2 & 4 & \cellcolor{customblue}21 & 30,891 & 2,048 & \cellcolor{customblue}21,433 & 64 & 128 & \cellcolor{customblue}64 \\
			& 1M & 0.31 & 0.10 & \cellcolor{customblue}0.31 & 7 & 3 & \cellcolor{customblue}{11~/~1} & 2 & 2 & \cellcolor{customblue}2 & 2,063 & 2,048 & \cellcolor{customblue}639 & 64 & 64 & \cellcolor{customblue}64 \\
\midrule
\multirow{3}{*}{\sf D6} & 100K & 0.59 & 0.85 & \cellcolor{customblue}0.96 & 5 & 10 & \cellcolor{customblue}{15~/~5} & 4 & 4 & \cellcolor{customblue}28 & 308 & 8,192 & \cellcolor{customblue}587 & 128 & 128 & \cellcolor{customblue}160 \\
			& 500K & 0.59 & 0.64 & \cellcolor{customblue}0.94 & 5 & 7 & \cellcolor{customblue}{23~/~3} & 3 & 3 & \cellcolor{customblue}16 & 354 & 2,048 & \cellcolor{customblue}578 & 96 & 96 & \cellcolor{customblue}96 \\
			& 1M & 0.51 & 0.34 & \cellcolor{customblue}0.73 & 8 & 3 & \cellcolor{customblue}{10~/~4} & 2 & 2 & \cellcolor{customblue}9 & 403 & 2,048 & \cellcolor{customblue}174 & 64 & 64 & \cellcolor{customblue}32 \\
\midrule
\multirow{3}{*}{\sf D7} & 100K & 0.87 & 0.86 & \cellcolor{customblue}0.99 & 5 & 10 & \cellcolor{customblue}{26~/~5} & 6 & 6 & \cellcolor{customblue}17 & 251 & 8,192 & \cellcolor{customblue}191 & 192 & 192 & \cellcolor{customblue}160 \\
			& 500K & 0.82 & 0.76 & \cellcolor{customblue}0.96 & 8 & 7 & \cellcolor{customblue}{10~/~6} & 4 & 3 & \cellcolor{customblue}10 & 495 & 2,048 & \cellcolor{customblue}106 & 128 & 96 & \cellcolor{customblue}64 \\
			& 1M & 0.47 & 0.41 & \cellcolor{customblue}0.76 & 5 & 3 & \cellcolor{customblue}{10~/~6} & 2 & 2 & \cellcolor{customblue}7 & 71 & 2,048 & \cellcolor{customblue}76 & 64 & 64 & \cellcolor{customblue}32 \\
\bottomrule
\end{tabular}
}
\vspace{-4pt}
\caption{Model performance vs. resource usage tested against Tofino1 switch (\SI{6.4}{Mbits} TCAM budget, 12 stages)~\cite{tofino}.}
\label{table:model-resources-format-5}
\vspace{-12pt}
\end{table*}

\begin{figure}[t]
    \centering
    \includegraphics[width=0.98\linewidth]{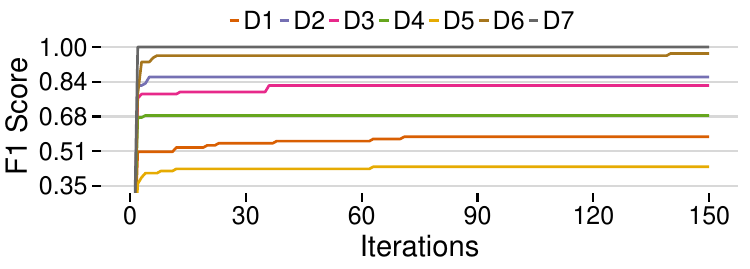}
    \vspace{-10pt}
    \caption{Number of BO search iterations to reach peak F1 score. All datasets converge within 150 iterations.}
    \label{fig:tta-iterations}
    \vspace{-18pt}
\end{figure}

We implement the \name{} framework in Python {\sf v3.10.13}, allowing seamless integration with widely used libraries such as Pandas~\cite{pandas}, Scikit-Learn~\cite{scikit_learn}, and TensorFlow~\cite{tensorflow}.
At the core of \name{}'s optimization process is HyperMapper~\cite{hypermapper} ({\sf commit:3dfa8a7}), which uses Bayesian Optimization (BO) to efficiently search for the best model configurations. 
Unlike other BO frameworks such as OpenTuner~\cite{opentuner}, HyperOpt~\cite{hyperopt}, and GPflow Opt~\cite{gpflow}, HyperMapper supports multi-objective optimization, diverse parameter types (real, integer, categorical), and feasibility testing~\cite{hypermapper}. 
These features allow us to optimize \name{} models for multiple goals simultaneously, such as maximizing F1 score and flow capacity. 
The feasibility testing feature also helps eliminate DT configurations that exceed available switch resources (\eg, TCAM limits) or introduce excessive resubmission traffic. 
The BO experiments are configured via YAML, specifying datasets, parameter space, objectives, and switch constraints.

To train and test \name{}'s partitioning strategy, we use the \textsf{DecisionTreeClassifier} class from Scikit-Learn~\cite{scikit_learn} and recursively generate subtrees for all partitions. 
We employ the Range Marking Algorithm, as described in NetBeacon~\cite{netbeacon}, to generate TCAM rules. 
We parallelize evaluations using Python's \textsf{ProcessPoolExecutor} from the \textsf{concurrent.futures} module to speed up the BO search. 
For data-plane inference, we implement \name{} in P4~\cite{p4-paper} and compile it using BF-SDE {\sf v9.11.0} for the Tofino1~\cite{tofino} switch. 
Finally, to install TCAM rules into the switch, we use \textsf{bfrt\_grpc} client API in Python.

In total, the \name{} framework consists of 4,700 lines of Python code. 
The P4 implementation and controller required 1,600 and 240 lines, respectively. 
Each experiment, corresponding to a dataset, was defined using 400 lines of YAML, totaling 2,800 lines for all seven datasets ({\sf D1--7}).

\section{Evaluation}
\label{sec:evaluation}

We evaluate \name{} DTs for their end-to-end classification performance and resource utilization (\S\ref{ssec:e2e}) and perform microbenchmark experiments (\S\ref{ssec:microbenchmarks}).

\subsection{Experiment Setup}
\label{ssec:exp-setup}

\noindent
\textbf{\em Testbed Environment.}
Our testbed includes two servers, one as a traffic generator and the other as a receiver, connected through an Edgecore Wedge 100-32X Tofino1 switch~\cite{tofino} running Stratum OS~\cite{stratum}.
Each server has a 64-core Intel Xeon Platinum 8358P CPU @\SI{2.60}{GHz}, \SI{512}{GB} RAM, and runs Proxmox VE {\sf v8.0.4}~\cite{proxmox}.
They are equipped with dual-port Intel XL710\cite{intel-xl710} 10/40G NICs and Nvidia ConnectX-6~\cite{mellanox-connect-x6-dx} 100G NICs, and use MoonGen~\cite{moongen} for traffic generation and reception.
The same servers run our Bayesian Optimization (BO) experiments, with 500 iterations and 16 parallel evaluations per iteration for each dataset.

\vspace{5pt}\noindent
\textbf{\em Baselines, and Real-World Applications and Environments.}
Our baselines include Leo~\cite{leo} and NetBeacon~\cite{netbeacon}, two state-of-the-art data plane DT implementations.
Both baselines support stateful (top-$k$) DTs and improve flow scalability with respect to tree depth via efficient TCAM rule generation.
To ensure fairness, we allocate the full switch pipeline (all stages) to each baseline, including ours, and evaluate the best-performing model each can support using all available hardware resources.
\Cref{table:datasets} summarizes the seven real-world datasets ({\sf D1--7}) used in our evaluation, spanning use cases such as intrusion detection, traffic classification, and modern attack detection (\eg, DoS, botnets, infiltration).
Together, these datasets allow for a robust evaluation of the performance of \name{} DTs against baselines across diverse network security tasks.
We also evaluate the volume of resubmitted traffic under two representative data center workloads---E1: Webserver (WS)~\cite{social_network}, with many long-lived flows, and E2: Hadoop (HD)~\cite{social_network}, characterized by short, bursty mice flows.

\vspace{3pt}\noindent
\textbf{\em Dataset Generation.}
To generate per-flow packet windows for training \name{} DTs, we extended the widely-used CICFlowMeter tool~\cite{cicflowmeter, cicflowmeter-paper-1, cicflowmeter-paper-2}.
By default, CICFlowMeter identifies flows by 5-tuples, computes 78 flow-level features, and emits statistics only at the FIN packet, discarding intermediate data.
We modified it to output statistics at every window boundary (\eg, each quarter of a flow for 4 partitions) and reset flow state after each window.
While NetBeacon's {\em phases} resemble \name{}'s windows, they differ in two key aspects: (a) phase intervals grow exponentially (\ie, 2, 4, 8, 16, ...), and (b) flow statistics are retained across phases, resulting in the same top-$k$ features being reused.
For each dataset in \Cref{table:datasets}, we generate up to 7 partitions\footnote{We experimented with more than 7 partitions, but classification accuracy significantly drops beyond this point.} for \name{}.
For NetBeacon~\cite{netbeacon}, we adopt the same phase intervals as their public artifact.

\subsection{End-to-End Analysis}
\label{ssec:e2e}

\noindent
\textbf{\em Pareto Frontier.}
Across all datasets, \name{} DTs consistently outperform the baselines by achieving a superior tradeoff---delivering higher accuracy at the same flow count.
This consistent ability to balance model performance (\ie, accuracy) with scalability (\ie, number of flows) allows \name{} to define the Pareto frontier across all evaluated datasets (\Cref{fig:pareto}).
The resulting tradeoff curves are monotonically decreasing: models yield higher accuracy when supporting fewer flows and progressively trade off feature coverage and model complexity to scale to larger flow counts.

\begin{table}[t]
\footnotesize

\begin{tabular}{p{27pt}rrrrrrr}
\toprule
{\bf Stages} & {\sf \textbf{D1}} & {\sf \textbf{D2}} & {\sf \textbf{D3}} & {\sf \textbf{D4}} & {\sf \textbf{D5}} & {\sf \textbf{D6}} & {\sf \textbf{D7}} \\
\toprule
{\sf Fetch} & 0.9s & 0.32s & 0.01s & 0.07s & 0.91s & 0.24s & 0.18s \\
\cellcolor{custompink}{\sf Training} & \cellcolor{custompink}556s & \cellcolor{custompink}228s & \cellcolor{custompink}10s & \cellcolor{custompink}84s & \cellcolor{custompink}725s & \cellcolor{custompink}163s & \cellcolor{custompink}111s \\
{\sf Optimizer} & 33s & 45s & 43s & 43s & 30s & 32s & 37s \\
{\sf Rulegen} & 0.8s & 0.99s & 0.91s & 1.08s & 0.71s & 0.71s & 0.91s \\
{\sf Backend} & 42$\mu$s & 45$\mu$s & 43$\mu$s & 42$\mu$s & 46$\mu$s & 44$\mu$s & 47$\mu$s \\
\midrule
{\sf Time} & 589s & 273s & 54s & 128s & 756s & 196s & 148s \\
\bottomrule
\end{tabular}
\vspace{-4pt}
\caption{Average time per iteration across different stages of the \namebold{} framework.}
\label{table:tta-breakdown}
\vspace{-6pt}
\end{table}

\begin{table}[t]
\centering
\footnotesize

\begin{tabular}{l lrrr}
\toprule
\multirow{2}{*}{\textbf{Environment}} & \multirow{2}{*}{\textbf{Data}} & \multicolumn{3}{c}{\textbf{Recirc. Bandwidth (Mbps)}} \\
\cmidrule(l){3-5}
 &  & \multicolumn{1}{c}{\textbf{100K}} & \multicolumn{1}{c}{\textbf{500K}} & \multicolumn{1}{c}{\textbf{1M}} \\
\toprule
\multirow{7}{*}{Webserver (WS)} 
 & {\sf D1} & 2.4$\pm$1.4 & 12.2$\pm$6.9 & 0.0$\pm$0.0 \\
 & {\sf D2} & 1.5$\pm$0.8 & 7.3$\pm$4.2 & 0.0$\pm$0.0 \\
 & {\sf D3} & 1.0$\pm$0.6 & 12.2$\pm$6.9 & 19.5$\pm$11.1 \\
 & {\sf D4} & 1.0$\pm$0.6 & 4.9$\pm$2.8 & 9.8$\pm$5.5 \\
 & {\sf D5} & 2.4$\pm$1.4 & 9.8$\pm$5.5 & 0.0$\pm$0.0 \\
 & {\sf D6} & 2.4$\pm$1.4 & 7.3$\pm$4.2 & 19.5$\pm$11.1 \\
 & {\sf D7} & 2.4$\pm$1.4 & 14.6$\pm$8.3 & 29.3$\pm$16.6 \\
\midrule
\multirow{7}{*}{Hadoop (HD)} 
 & {\sf D1} & 5.0$\pm$2.0 & 25.0$\pm$9.9 & 0.0$\pm$0.0 \\
 & {\sf D2} & 3.0$\pm$1.2 & 15.0$\pm$6.0 & 0.0$\pm$0.0 \\
 & {\sf D3} & 2.0$\pm$0.8 & 25.0$\pm$9.9 & 40.0$\pm$15.9 \\
 & {\sf D4} & 2.0$\pm$0.8 & 10.0$\pm$4.0 & 20.0$\pm$8.0 \\
 & {\sf D5} & 5.0$\pm$2.0 & 20.0$\pm$8.0 & 0.0$\pm$0.0 \\
 & {\sf D6} & 5.0$\pm$2.0 & 15.0$\pm$6.0 & 40.0$\pm$15.9 \\
 & {\sf D7} & 5.0$\pm$2.0 & 30.0$\pm$11.9 & 60.0$\pm$23.9 \\
\bottomrule
\end{tabular}
\vspace{-4pt}
\caption{Maximum recirculation bandwidth (Mbps) of \namebold{} partitioned trees when processing datasets ({\sf \textbf{D1--7}}) for the two datacenter environments, Webserver (WS) and Hadoop (HD), with varying flow sizes.}
\label{table:bm-recirculation-bw}
\vspace{-15pt}

\end{table}

\begin{figure}[t]
    \centering
    \includegraphics[width=1.0\linewidth]{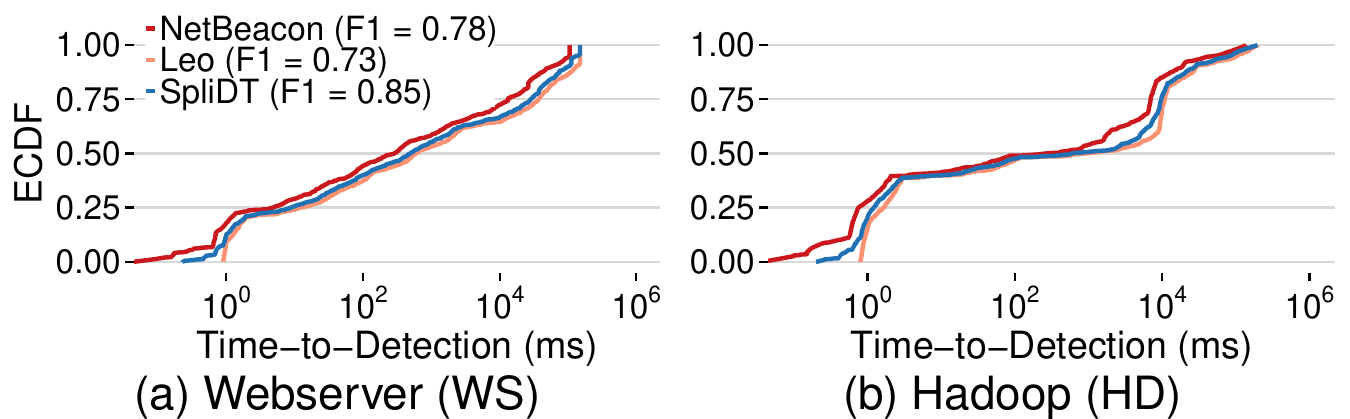}
    \vspace{-18pt}
    \caption{Time-to-detection (TTD) of {\sf \textbf{D3}} for environments: WS and HD. Other datasets show a similar trend.}
    \label{fig:ttd-cdf}
    \vspace{5pt}
    \centering
    \includegraphics[width=1.0\linewidth]{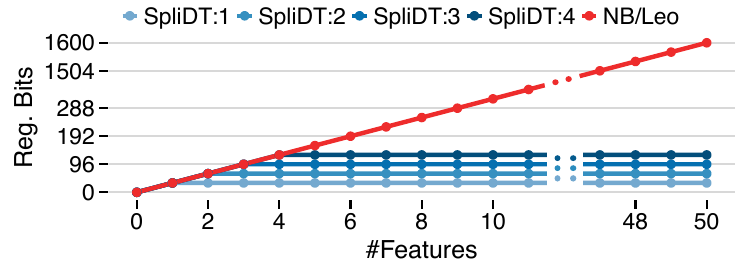}
    \vspace{-20pt}
    \caption{Register sizes (in bits) versus number of features supported by each model. \namebold{}:{\em k} is a partitioned tree with {\em k} features per subtree.}
    \label{fig:register-size}
	\vspace{5pt}
    \centering
    \includegraphics[width=1.0\linewidth]{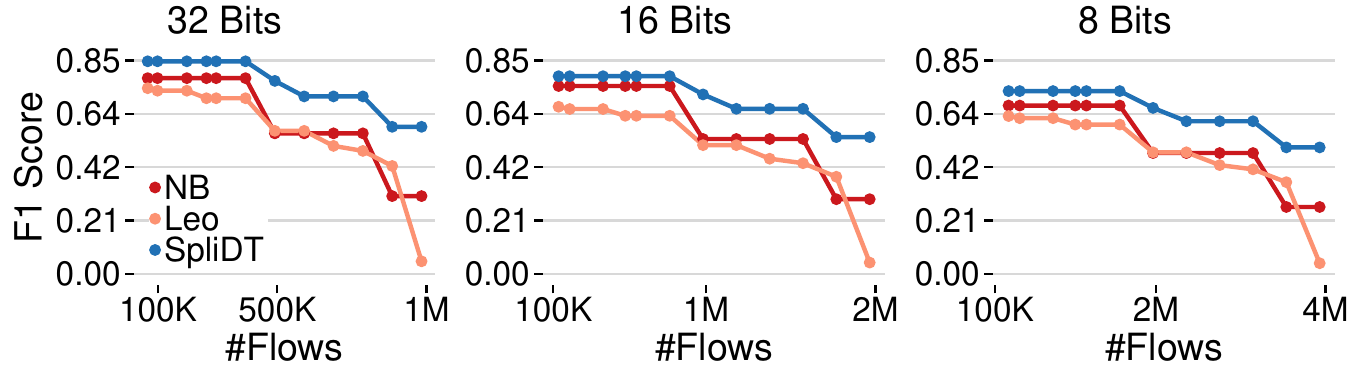}
    \vspace{-20pt}
    \caption{Pareto frontier of {\sf \textbf{D3}} versus bit precisions.}
    \label{fig:bit-precision}
	\vspace{-22pt}
\end{figure}

\vspace{5pt}\noindent
{\em \textbf{Feature Scalability and Resource Utilization} (TCAMs, Registers, and Recirculation Bandwidth)}.
As shown in \Cref{table:model-resources-format-5}, \name{} DTs consistently deliver the highest accuracy across datasets, balancing tree depth and partitioning effectively.
They use less register space while maintaining competitive accuracy, and optimize TCAM usage with manageable register overhead.
For instance, in {\sf D6}, \name{} supports more features than both baselines across all flow sizes (within 160-, 96-, and 32-bit budgets for 100K, 500K, and 1M flows, respectively), while also deploying deeper trees.
As shown in \Cref{table:bm-recirculation-bw} and \Cref{fig:ttd-cdf}, \name{}’s packet recirculation remains well within the 100 Gbps resubmission buffer capacity and does not degrade model responsiveness, matching NetBeacon~\cite{netbeacon} and Leo~\cite{leo} in per-flow time-to-detection (TTD).
These results underscore the efficiency of \name{} DTs in maximizing accuracy under the same resource constraints as competing baselines.

\vspace{5pt}\noindent
\textbf{\em Offline Software Overhead.}
\name{}'s offline design search reaches maximum accuracy for all datasets within 150 iterations (\Cref{fig:tta-iterations}).
As shown in \Cref{table:tta-breakdown}, training dominates the per-iteration cost (88\% on average), followed by the BO stage (12\%), which leverages HyperMapper~\cite{hypermapper} to guide the search toward high-performing models.
Training is computationally intensive due to the recursive nature of \name{} DTs, which involves deeper trees, more subtrees, and repeated dataset partitioning.
Despite this, the full search completes in an average of 3.8 hours per dataset, yielding optimized, deployment-ready models.

\begin{figure*}[t]
    \centering
    \includegraphics[width=1.0\linewidth]{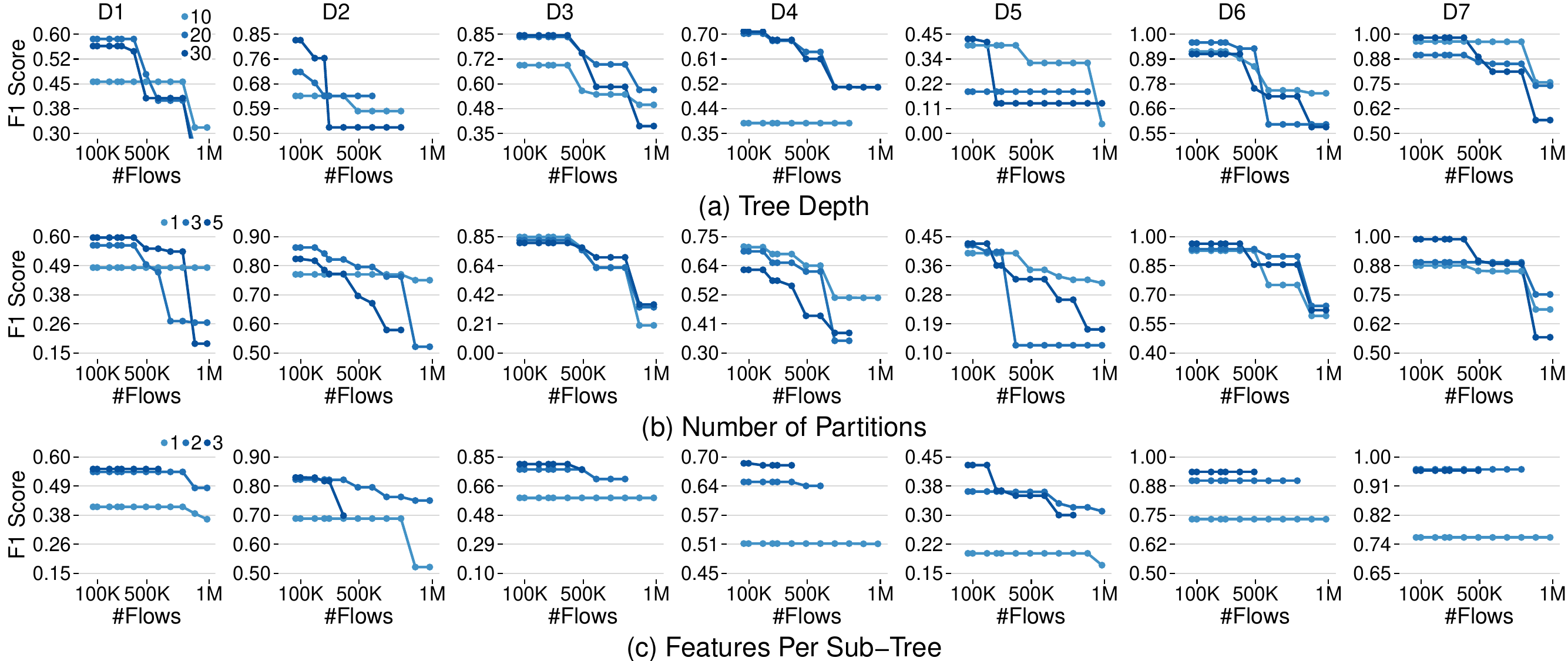}
    \vspace{-15pt}
    \caption{Pareto frontiers for \namebold{} partitioned trees under varying constraints (top to bottom): (a) fixed tree depth, (b) fixed number of partitions, and (c) fixed number of features per subtree.}
    \label{fig:bm-pareto}
	\vspace{5pt}
    \centering
    \includegraphics[width=1.0\linewidth]{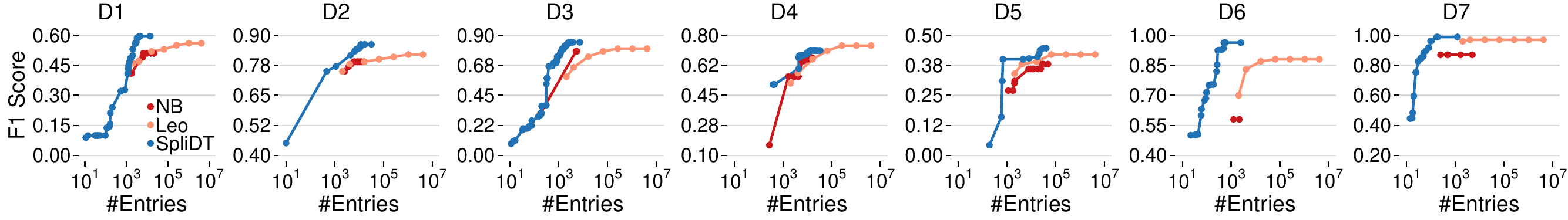}
    \vspace{-15pt}
    \caption{Comparison of \#TCAM entries against F1 score for \namebold{} versus baselines.}
    \label{fig:bm-tcam-f1-score}
    \vspace{-15pt}
\end{figure*}

\subsection{Microbenchmarks}
\label{ssec:microbenchmarks}

\noindent
\textbf{\em \namebold{} framework finds optimal model parameters for partitioned decision trees.}
The \name{} DSE framework navigates the parameter space for tree partitioning to identify high-performing models for a given number of flows.
\Cref{fig:bm-pareto}a shows the Pareto frontiers for fixed tree depths (10, 20, 30) as we vary the features per subtree and partition count. 
For instance, in {\sf D2}, a depth of 30 generally yields a better frontier than depth 10---except at high flow counts (\eg, 1M flows), where the simpler model performs comparably.
Each dataset benefits from a different tree depth depending on its feature complexity and class distribution, and \name{} effectively adapts to these differences.
\Cref{fig:bm-pareto}b fixes the number of partitions (1, 3, 5), showing that fewer partitions often yield better frontiers due to more packets per window, enhancing model accuracy.
Lastly, \Cref{fig:bm-pareto}c varies the number of features per subtree (1, 2, 3): using more features improves accuracy but reduces scalability; fewer features increase flow support at the cost of classification performance.

\vspace{5pt}\noindent
\textbf{\em \namebold{} DTs outperform baselines for any given TCAM budget.}
\Cref{fig:bm-tcam-f1-score} shows that \name{} DTs consistently achieve higher accuracy than the baselines across different TCAM budgets.
This improvement stems from reducing the match key size by decreasing the number of features ($k$) used in the model table for the active subtree.

\vspace{5pt}\noindent
\textbf{\em \namebold{} DTs recirculate limited traffic without impacting end-to-end flow-level time-to-detection (TTD).}
\Cref{table:bm-recirculation-bw} shows that in both environments: WS and HD, packet recirculation by \name{} DTs stays well within the resubmission buffer capacity of \SI{100}{Gbps}.
Additionally, as shown in \Cref{fig:ttd-cdf}, the per-flow time-to-detection (TTD) for {\sf D3}---defined as the time from the start of tree traversal to the final inference decision---closely matches that of NetBeacon and Leo, while sustaining 9\% and 16\% higher F1 score, respectively.
This indicates that recirculation overhead in \name{} DTs does not negatively impact model performance or responsiveness.

\vspace{5pt}\noindent
\textbf{\em \namebold{} DTs scale feature count with constant register space.}
\name{} DTs maintain a constant register footprint, based solely on $k$ features per subtree, regardless of the total number of unique features used across the entire tree (\Cref{fig:register-size}).
As shown in \Cref{table:model-resources-format-5}, \name{} can store up to 30 distinct 32-bit features within a 128-bit register budget.
This demonstrates that by dynamically multiplexing features across partitions and subtrees, \name{} minimizes register overhead.
This efficient use of stateful memory enables the deployment of deeper, more complex DTs in the data plane.

\vspace{5pt}\noindent
\textbf{\em \namebold{} DTs can accommodate a larger number of flows with reduced feature bit precision while maintaining higher F1 scores.}
We lower the bit precision of all features from 32 bits to 16 and 8 bits, respectively, and measure the resulting impact on accuracy and flow scalability. 
\Cref{fig:bit-precision} shows that this reduction affects all models similarly, as they are all decision trees. 
On average, we observe a 7\% (14\%) drop in accuracy with 16-bit (8-bit) precision, while the maximum number of supported flows increases to 2M (4M). 
In both cases, \name{} continues to yield a better Pareto frontier than the baselines, due to its enhanced feature coverage.

\section{Limitations \&  Future Work}
\label{sec:limitations}

\noindent
\textbf{\em Adaptive Window Sizing.}
Currently, \name{} uses fixed per-flow window sizes, which may not always reflect real-world traffic dynamics. 
This limitation can affect both model accuracy and resource efficiency, particularly under bursty or highly variable traffic patterns. 
Future work could explore adaptive window sizing across partitions, enabling \name{} to optimize feature extraction based on observed flow characteristics.

\vspace{5pt}
\noindent
\textbf{\em Security \& Robustness.}
\name{}'s reliance on flow size information stored in packet headers makes it vulnerable to scenarios where this information is hacked or spoofed. 
If an attacker manipulates this field, it could lead to incorrect window boundaries, resulting in misclassifications, resource mismanagement, or even denial-of-service (DoS) attacks. 
Enhancing security measures to validate and protect header information is a critical area for future improvement.

\section{Related Work}
\label{sec:related}

\noindent
\textbf{\em Decision Tree (DT)-Based Inference in the Data Plane.}
Several efforts have mapped decision trees (DTs) to programmable data planes (\ie, switches) for real-time classification. 
Leo~\cite{leo} scales DTs by optimizing match-action table (MAT) representations, while NetBeacon~\cite{netbeacon} introduces ternary table encodings for large-scale inference. 
Mousika~\cite{mousika} applies knowledge distillation to optimize binary DTs, reducing resource usage while preserving hardware compatibility. 
Planter~\cite{planter} incorporates packet-level summaries into DT-based classification, and IIsy~\cite{iisy-hotnets} maps ML models to MAT pipelines to improve inference efficiency. 
pForest~\cite{pforest} generalizes DTs to random forests, enabling dynamic feature selection based on real-time traffic. 
Unlike these approaches, \name{} removes the static top-$k$ feature constraint and enables dynamic feature allocation across tree partitions. 
In contrast to prior single-pass DT execution, \name{} introduces incremental inference via subtree transitions---optimizing stateful feature storage and scaling to millions of flows with reduced TCAM overhead.

\vspace{2pt}\noindent
\textbf{\em Neural Network (NN)-Based Inference in the Data Plane.}
NN-based approaches in the data plane prioritize high expressiveness but often encounter resource limitations.
Taurus~\cite{taurus} proposes a MapReduce-inspired ML inference framework, while Homunculus~\cite{homunculus} automates ML model deployment on switches and SmartNICs.
Brain-on-Switch (BoS)\cite{bos} incorporates RNNs and transformers for sequential processing, improving accuracy but demanding significant compute resources.
N3IC\cite{n3ic} explores binary neural networks for low-latency inference, and ServeFlow~\cite{serveflow} introduces neural accelerators to support high-accuracy classification with minimal overhead.
AC-DC~\cite{acdc} aims to balance performance and efficiency in dynamic traffic scenarios.
To support such workloads, recent converged platforms like NVIDIA BlueField-3~\cite{nvidia-bluefield-dpu} combine embedded CPU cores, network interfaces, and hardware accelerators within a SmartNIC, enabling ML inference offload without host CPU intervention.
Similarly, Broadcom's Trident 5~\cite{broadcom-trident5} introduces a rigid on-chip neural inference engine (NetGNT) that performs real-time traffic analysis at line rate without impacting switch throughput or latency.
While such architectures are promising for NN-based inference, they rely on dedicated hardware engines and are inherently constrained in terms of general-purpose model flexibility and resource sharing.
In contrast, \name{} is designed explicitly for resource-constrained, line-rate execution in existing data planes.
By partitioning decision trees and reusing MAT stages and registers via recirculation, \name{} supports scalable, high-accuracy inference without requiring external accelerators or specialized NN engines.


\vspace{2pt}\noindent
\textbf{\em Lookaside Accelerators for ML inference.}
Modern ML inference increasingly leverages high-performance accelerators such as GPUs~\cite{nvidia-t4}, FPGAs~\cite{xilinx-sn-1000, xilinx-u250}, TPUs~\cite{google-cloud-tpu}, and SmartNICs~\cite{amd-pensando, nvidia-bluefield-dpu}.
These platforms excel at offline or near-real-time inference, where flexible compute and memory resources enable support for deep neural networks and large-scale models.
As a result, many cloud and data center architectures adopt a {\em lookaside} model, offloading packets or flow metadata to external co-processors for ML inference~\cite{sonata, gpu, dynamap}.
While this approach works well for sampled traffic or background analytics, it is ill-suited for high-speed networks that demand line-rate packet processing with microsecond-level latency~\cite{netbeacon, leo, broadcom-trident5}.
In such environments, external accelerators cannot consistently match the throughput and latency requirements of the data plane, forcing operators to rely on packet sampling, which reduces coverage and weakens real-time detection guarantees.
In contrast, \name{} is explicitly designed for in-network inference, enabling scalable, line-rate DT processing entirely within the programmable data plane.
By eliminating the need for lookaside offload and avoiding dependence on specialized external accelerators, \name{} ensures that line-rate, low-latency inference can be performed directly on the switch fast path---making it a practical solution for modern, high-throughput network environments.

\section{Conclusion}
\label{sec:conclusion}

We demonstrate in this paper how \name{} recasts decision tree (DT) deployment in programmable data planes through partitioned, window-based inference. 
Unlike prior approaches that rely on static top-$k$ feature sets and one-shot execution, \name{} enables incremental inference via subtree transitions and dynamic feature reuse.
Combined with a Bayesian optimization (BO)-based training pipeline, \name{} balances model accuracy and hardware efficiency, scaling to millions of flows at line rate.


\name{} shows that complex and expressive in-network ML inference is not only feasible but practical---without compromising model accuracy or performance.
As network traffic grows in volume and speed, \name{} offers a scalable and efficient path to real-time, high-throughput inference under tight hardware constraints.
By enabling informed in-network decision-making, \name{} contributes to ongoing efforts toward a more responsible use of ML in networking.

\label{lastpage}

\if\showacks1
    \section*{Acknowledgements}
To Robert, for the bagels and explaining CMYK and color spaces.

\fi

{\small
\bibliographystyle{plain}
\bibliography{bibs/paper}}

\begin{thebibliography}{10}

\bibitem{amd-pensando}
AMD.
\newblock {Pensando}.
\newblock \url{https://www.amd.com/en/accelerators/pensando}, last accessed: 06/05/2025.

\bibitem{malware_traffic_classification}
Blake Anderson and David McGrew.
\newblock {Machine Learning for Encrypted Malware Traffic Classification: Accounting for Noisy Labels and Non-Stationarity}.
\newblock In {\em ACM SIGKDD International Conference on Knowledge Discovery and Data Mining}, 2017.

\bibitem{opentuner}
Jason Ansel, Shoaib Kamil, Kalyan Veeramachaneni, Jonathan Ragan-Kelley, Jeffrey Bosboom, Una-May O'Reilly, and Saman Amarasinghe.
\newblock {Opentuner: An Extensible Framework for Program Autotuning}.
\newblock In {\em Proceedings of the 23rd International Conference on Parallel Architectures and Compilation}, 2014.

\bibitem{ndss_traffic_analysis}
Alireza Bahramali, Amir Houmansadr, Ramin Soltani, Dennis Goeckel, and Don Towsley.
\newblock {Practical Traffic Analysis Attacks on Secure Messaging Applications}.
\newblock In {\em NDSS}, 2020.

\bibitem{flowlens}
Diogo Barradas, Nuno Santos, Luis Rodrigues, Salvatore Signorello, Fernando M.~V. Ramos, and André Madeira.
\newblock {FlowLens: Enabling Efficient Flow Classification for ML-based Network Security Applications}.
\newblock In {\em NDSS}, 2021.

\bibitem{understanding-datacenter}
Theophilus Benson, Ashok Anand, Aditya Akella, and Ming Zhang.
\newblock {Understanding data center traffic characteristics}.
\newblock {\em ACM SIGCOMM Computer Communication Review (CCR)}, 2010.

\bibitem{hyperopt}
J.~Bergstra, D.~Yamins, and D.~D. Cox.
\newblock {Making a Science of Model Search: Hyperparameter Optimization in Hundreds of Dimensions for Vision Architectures}.
\newblock In {\em ICML}, 2013.

\bibitem{p4-paper}
Pat Bosshart, Dan Daly, Glen Gibb, Martin Izzard, Nick McKeown, Jennifer Rexford, Cole Schlesinger, Dan Talayco, Amin Vahdat, George Varghese, and David Walker.
\newblock {P4: Programming Protocol-Independent Packet Processors}.
\newblock In {\em ACM SIGCOMM Computer Communication Review (CCR)}, 2014.

\bibitem{rmt}
Pat Bosshart, Glen Gibb, Hun-Seok Kim, George Varghese, Nick McKeown, Martin Izzard, Fernando Mujica, and Mark Horowitz.
\newblock {Forwarding Metamorphosis: Fast Programmable Match-Action Processing in Hardware for SDN}.
\newblock In {\em ACM SIGCOMM}, 2013.

\bibitem{broadcom-trident5}
BROADCOM.
\newblock {Trident 5 / BCM78800 Series}.
\newblock \url{https://www.broadcom.com/products/ethernet-connectivity/switching/strataxgs/bcm78800}, last accessed: 06/05/2025.

\bibitem{broadcom-trident4}
BROADCOM.
\newblock {Trident4/BCM56880 Series}.
\newblock \url{https://www.broadcom.com/products/ethernet-connectivity/switching/strataxgs/bcm56880-series}, last accessed: 06/05/2025.

\bibitem{pforest}
Coralie Busse-Grawitz, Roland Meier, Alexander Dietmüller, Tobias Bühler, and Laurent Vanbever.
\newblock {pForest: In-Network Inference with Random Forests}.
\newblock {\em arXiv preprint arXiv:1909.05680}, 2022.

\bibitem{cicids17}
{Canadian Institute for Cybersecurity}.
\newblock {CIC IDS 2017 Dataset}.
\newblock \url{https://www.unb.ca/cic/datasets/ids-2017.html}, last accessed: 06/05/2025.

\bibitem{cicids18}
{Canadian Institute for Cybersecurity}.
\newblock {CIC IDS 2018 Dataset}.
\newblock \url{https://www.unb.ca/cic/datasets/ids-2018.html}, last accessed: 06/05/2025.

\bibitem{ciciomt24}
{Canadian Institute for Cybersecurity}.
\newblock {CIC IoMT 2024 Dataset}.
\newblock \url{https://www.unb.ca/cic/datasets/iomt-dataset-2024.html}, last accessed: 06/05/2025.

\bibitem{ciciot23}
{Canadian Institute for Cybersecurity}.
\newblock {CIC IoT 2023 Dataset}.
\newblock \url{https://www.unb.ca/cic/datasets/iotdataset-2023.html}, last accessed: 06/05/2025.

\bibitem{cicvpn}
{Canadian Institute for Cybersecurity}.
\newblock {CIC VPN Dataset}.
\newblock \url{https://www.unb.ca/cic/datasets/vpn.html}, last accessed: 06/05/2025.

\bibitem{google-cloud-tpu}
Google Cloud.
\newblock {Tensor Processing Units (TPUs)}.
\newblock \url{https://cloud.google.com/tpu}, 2025.

\bibitem{intel-p4-insight}
Intel Corporation.
\newblock {Intel P4 Insight}.
\newblock \url{https://p4.org/onf-product/intel-p4-insight/}, last accessed: 06/05/2025.

\bibitem{intel-p4studio}
Intel Corporation.
\newblock {Intel® P4 Studio}.
\newblock \url{https://www.intel.com/content/www/us/en/products/details/network-io/intelligent-fabric-processors/p4-studio.html}, last accessed: 06/05/2025.

\bibitem{nvidia-t4}
NVIDIA Corporation.
\newblock {NVIDIA T4 Tensor Core GPU}.
\newblock \url{https://www.nvidia.com/en-us/data-center/tesla-t4/}, 2025.

\bibitem{artifact}
{Details omitted for double-blind review}.

\bibitem{pcc-congestion}
Mo~Dong, Qingxi Li, Doron Zarchy, P.~Brighten Godfrey, and Michael Schapira.
\newblock {PCC: Re-Architecting Congestion Control for Consistent High Performance}.
\newblock In {\em USENIX NSDI}, 2015.

\bibitem{horus}
Yutao Dong, Qing Li, Kaidong Wu, Ruoyu Li, Dan Zhao, Gareth Tyson, Junkun Peng, Yong Jiang, Shutao Xia, and Mingwei Xu.
\newblock {{HorusEye}: A Realtime {IoT} Malicious Traffic Detection Framework using Programmable Switches}.
\newblock In {\em USENIX Security}, 2023.

\bibitem{lucid}
R.~Doriguzzi-Corin, S.~Millar, S.~Scott-Hayward, J.~Martínez-del Rincón, and D.~Siracusa.
\newblock {Lucid: A Practical, Lightweight Deep Learning Solution for DDoS Attack Detection}.
\newblock {\em IEEE Transactions on Network and Service Management}, 2020.

\bibitem{moongen}
Paul Emmerich, Sebastian Gallenm{\"u}ller, Daniel Raumer, Florian Wohlfart, and Georg Carle.
\newblock {Moongen: A Scriptable High-Speed Packet Generator}.
\newblock In {\em ACM IMC}, 2015.

\bibitem{bohb}
Stefan Falkner, Aaron Klein, and Frank Hutter.
\newblock {BOHB: Robust and efficient hyperparameter optimization at scale}.
\newblock In {\em ICML}, 2018.

\bibitem{stratum}
Open~Networking Foundation.
\newblock {Stratum OS}.
\newblock \url{https://www.opennetworking.org/stratum/}, last accessed: 06/05/2025.

\bibitem{Fu0023}
Chuanpu Fu, Qi~Li 0002, and Ke~Xu 0002.
\newblock {Detecting Unknown Encrypted Malicious Traffic in Real Time via Flow Interaction Graph Analysis}.
\newblock In {\em NDSS}, 2023.

\bibitem{npl}
Saikrishna Garlapati.
\newblock {Network Programming Language (NPL) Specification}.
\newblock \href{https://www.scribd.com/document/430082948/Network-programming-Language-NPL}{https://www.scribd.com/document/430082948/Network-programming-Language-NPL}, last accessed: 06/05/2025.

\bibitem{cicflowmeter-paper-1}
Gerard~Drapper Gil, Arash~Habibi Lashkari, Mohammad Mamun, and Ali~A Ghorbani.
\newblock {Characterization of encrypted and VPN traffic using time-related features}.
\newblock In {\em Proceedings of the 2nd international conference on information systems security and privacy}, 2016.

\bibitem{cicflowmeter}
Arash Habibi~Lashkari (GitHub).
\newblock {CICFlowMeter}.
\newblock \url{https://github.com/ahlashkari/CICFlowMeter/tree/master}, last accessed: 06/05/2025.

\bibitem{postgresql}
PostgreSQL Global~Development Group.
\newblock {PostgreSQL}.
\newblock \url{https://www.postgresql.org}, last accessed: 06/05/2025.

\bibitem{sonata}
A.~Gupta, R.~Harrison, M.~Canini, N.~Feamster, J.~Rexford, and W.~Willinger.
\newblock {Sonata: Query-driven network telemetry}.
\newblock In {\em ACM SIGCOMM}, 2018.

\bibitem{netfound}
Satyandra Guthula, Roman Beltiukov, Navya Battula, Wenbo Guo, Arpit Gupta, and Inder Monga.
\newblock {netFound: Foundation Model for Network Security}.
\newblock {\em arXiv preprint arXiv:2310.17025}, 2025.

\bibitem{cubic-tcp}
Sangtae Ha, Injong Rhee, and Lisong Xu.
\newblock {CUBIC: A New TCP-Friendly High-Speed TCP Variant}.
\newblock {\em ACM SIGOPS Operating Systems Review}, 2008.

\bibitem{ndp}
Mark Handley, Costin Raiciu, Alexandru Agache, Andrei Voinescu, Andrew~W. Moore, Gianni Antichi, and Marcin W\'{o}jcik.
\newblock {Re-Architecting Datacenter Networks and Stacks for Low Latency and High Performance}.
\newblock In {\em ACM SIGCOMM}, 2017.

\bibitem{crc32}
{He3 Team}.
\newblock {{Understanding the CRC32 Hash: A Comprehensive Guide}}.
\newblock \href{https://he3.app/blogs/understanding-the-crc32-hash-a-comprehensive-guide/}{https://he3.app/blogs/understanding-the-crc32-hash-a-comprehensive-guide/}, last accessed: 06/05/2025.

\bibitem{mongodb}
MongoDB Inc.
\newblock {MongoDB}.
\newblock \url{https://www.mongodb.com}, last accessed: 06/05/2025.

\bibitem{intel-xl710}
Intel.
\newblock {Intel Ethernet Network Adapter X710}.
\newblock \url{https://www.intel.com/content/www/us/en/products/details/ethernet/700-network-adapters/x710-network-adapters/products.html}, last accessed: 06/05/2025.

\bibitem{tofino}
Intel.
\newblock {Tofino: P4-programmable Ethernet switch ASIC that delivers better performance at lower power}.
\newblock \href{https://www.intel.com/content/www/us/en/products/network-io/programmable-ethernet-switch/tofino-series.html}{https://www.intel.com/content/www/us/en/products/network-io/programmable-ethernet-switch/tofino-series.html}, last accessed: 06/05/2025.

\bibitem{tofino2}
Intel.
\newblock {Tofino2: Second-generation P4-programmable Ethernet Switch ASIC that Continues to Deliver Programmability without Compromise}.
\newblock \href{https://www.intel.com/content/www/us/en/products/network-io/programmable-ethernet-switch/tofino-2-series.html}{https://www.intel.com/content/www/us/en/products/network-io/programmable-ethernet-switch/tofino-2-series.html}, last accessed: 06/05/2025.

\bibitem{leo}
Syed~Usman Jafri, Sanjay Rao, Vishal Shrivastav, and Mohit Tawarmalani.
\newblock {Leo: Online {ML-based} Traffic Classification at {Multi-Terabit} Line Rate}.
\newblock In {\em USENIX NSDI}, 2024.

\bibitem{acdc}
Xi~Jiang, Shinan Liu, Saloua Naama, Francesco Bronzino, Paul Schmitt, and Nick Feamster.
\newblock {AC-DC: Adaptive Ensemble Classification for Network Traffic Identification}.
\newblock {\em arXiv preprint arXiv:2302.11718}, 2023.

\bibitem{gpflow}
Nicolas Knudde, Joachim van~der Herten, Tom Dhaene, and Ivo Couckuyt.
\newblock {GPflowOpt: A Bayesian Optimization Library Using TensorFlow}.
\newblock {\em arXiv preprint arXiv:1711.03845}, 2017.

\bibitem{intel-ipu-offload}
Patricia Kummrow.
\newblock {The IPU: A New, Strategic Resource for Cloud Service Providers}.
\newblock \href{https://community.intel.com/t5/Blogs/Tech-Innovation/Data-Center/The-IPU-A-New-Strategic-Resource-for-Cloud-Service-Providers/post/1335081}{https://community.intel.com/t5/Blogs/Tech-Innovation/Data-Center/The-IPU-A-New-Strategic-Resource-for-Cloud-Service-Providers/post/1335081}, last accessed: 06/05/2025.

\bibitem{cicflowmeter-paper-2}
Arash~Habibi Lashkari, Gerard~Draper Gil, Mohammad Saiful~Islam Mamun, and Ali~A Ghorbani.
\newblock {Characterization of tor traffic using time based features}.
\newblock In {\em International Conference on Information Systems Security and Privacy}, 2017.

\bibitem{Li:2019:HHP:3341302.3342085}
Yuliang Li, Rui Miao, Hongqiang~Harry Liu, Yan Zhuang, Fei Feng, Lingbo Tang, Zheng Cao, Ming Zhang, Frank Kelly, Mohammad Alizadeh, and Minlan Yu.
\newblock {HPCC: High Precision Congestion Control}.
\newblock In {\em ACM SIGCOMM}, 2019.

\bibitem{smac3}
Marius Lindauer, Katharina Eggensperger, Matthias Feurer, Andr{\'e} Biedenkapp, Difan Deng, Carolin Benjamins, Tim Ruhkopf, Ren{\'e} Sass, and Frank Hutter.
\newblock {SMAC3: A versatile Bayesian optimization package for hyperparameter optimization}.
\newblock {\em Journal of Machine Learning Research (JMLR)}, 2022.

\bibitem{serveflow}
Shinan Liu, Ted Shaowang, Gerry Wan, Jeewon Chae, Jonatas Marques, Sanjay Krishnan, and Nick Feamster.
\newblock {ServeFlow: A Fast-Slow Model Architecture for Network Traffic Analysis}.
\newblock {\em arXiv preprint arXiv:2402.03694}, 2024.

\bibitem{mao2017neural}
Hongzi Mao, Ravi Netravali, and Mohammad Alizadeh.
\newblock {Neural Adaptive Video Streaming with Pensieve}.
\newblock In {\em ACM SIGCOMM}, 2017.

\bibitem{homa}
Behnam Montazeri, Yilong Li, Mohammad Alizadeh, and John Ousterhout.
\newblock {Homa: A Receiver-Driven Low-Latency Transport Protocol Using Network Priorities}.
\newblock In {\em ACM SIGCOMM}, 2018.

\bibitem{hypermapper}
Luigi Nardi, Bruno Bodin, Sajad Saeedi, Emanuele Vespa, Andrew~J Davison, and Paul~HJ Kelly.
\newblock {Algorithmic Performance-accuracy Trade-off in 3D Vision Applications using Hypermapper}.
\newblock In {\em IEEE IPDPSW}, 2017.

\bibitem{mellanox-connect-x6-dx}
Nvidia.
\newblock {ConnectX-6 Network Adapters}.
\newblock \href{https://www.nvidia.com/en-in/networking/ethernet/connectx-6-dx/}{https://www.nvidia.com/en-us/networking/ethernet/connectx-6-dx/}, last accessed: 06/05/2025.

\bibitem{doca}
Nvidia.
\newblock {DOCA Documentation}.
\newblock \url{https://docs.nvidia.com/doca/archive/2-9-2/doca+p4+developer+tools/index.html}, last accessed: 06/05/2025.

\bibitem{nvidia-bluefield-dpu}
Nvidia.
\newblock {Nvidia BlueField Data Processing Units}.
\newblock \url{https://www.nvidia.com/en-us/networking/products/data-processing-unit/}, last accessed: 06/05/2025.

\bibitem{nvidia-spectrumx}
{NVIDIA Corporation}.
\newblock {NVIDIA Spectrum-X: Ethernet Networking Platform for AI}.
\newblock \url{https://www.nvidia.com/en-us/networking/spectrumx/}, last accessed: 06/05/2025.

\bibitem{pandas}
pandas.
\newblock {pandas}.
\newblock \url{https://pandas.pydata.org/}, last accessed: 06/05/2025.

\bibitem{scikit_learn}
Fabian Pedregosa, Ga{\"e}l Varoquaux, Alexandre Gramfort, Vincent Michel, Bertrand Thirion, Olivier Grisel, Mathieu Blondel, Peter Prettenhofer, Ron Weiss, Vincent Dubourg, et~al.
\newblock {Scikit-learn: Machine learning in Python}.
\newblock {\em Journal of machine learning research}, 2011.

\bibitem{proxmox}
Proxmox.
\newblock {Proxmox}.
\newblock \href{https://www.proxmox.com/en/}{https://www.proxmox.com/en/}, last accessed: 06/05/2025.

\bibitem{social_network}
Arjun Roy, Hongyi Zeng, Jasmeet Bagga, George Porter, and Alex~C. Snoeren.
\newblock {Inside the Social Network's (Datacenter) Network}.
\newblock In {\em ACM SIGCOMM}, 2015.

\bibitem{eRSS}
Alexander Rucker, Muhammad Shahbaz, Tushar Swamy, and Kunle Olukotun.
\newblock {Elastic RSS: Co-Scheduling Packets and Cores Using Programmable NICs}.
\newblock In {\em APNet}, 2019.

\bibitem{netasm}
Muhammad Shahbaz and Nick Feamster.
\newblock {The case for an intermediate representation for programmable data planes}.
\newblock In {\em SOSR}, 2015.

\bibitem{dynamap}
Chaofan Shou, Rohan Bhatia, Arpit Gupta, Rob Harrison, Daniel Lokshtanov, and Walter Willinger.
\newblock {Query planning for robust and scalable hybrid network telemetry systems}.
\newblock {\em Proceedings of the ACM on Networking}, 2024.

\bibitem{hyper_tuning}
Sebastian Simon, Nikolay Kolyada, Christopher Akiki, Martin Potthast, Benno Stein, and Norbert Siegmund.
\newblock {Exploring Hyperparameter Usage and Tuning in Machine Learning Research}.
\newblock In {\em IEEE/ACM 2nd International Conference on AI Engineering--Software Engineering for AI (CAIN)}, 2023.

\bibitem{n3ic}
Giuseppe Siracusano, Salvator Galea, Davide Sanvito, Mohammad Malekzadeh, Gianni Antichi, Paolo Costa, Hamed Haddadi, and Roberto Bifulco.
\newblock {Re-architecting Traffic Analysis with Neural Network Interface Cards}.
\newblock In {\em USENIX NSDI}, 2022.

\bibitem{taurus}
Tushar Swamy, Alexander Rucker, Muhammad Shahbaz, Ishan Gaur, and Kunle Olukotun.
\newblock {Taurus: A Data Plane Architecture for Per-Packet ML}.
\newblock In {\em ASPLOS}, 2022.

\bibitem{homunculus}
Tushar Swamy, Annus Zulfiqar, Luigi Nardi, Muhammad Shahbaz, and Kunle Olukotun.
\newblock {Homunculus: Auto-Generating Efficient Data-Plane ML Pipelines for Datacenter Networks}.
\newblock In {\em ASPLOS}, 2023.

\bibitem{tensorflow}
Tensorflow.
\newblock {Tensorflow}.
\newblock \url{https://www.tensorflow.org/}, last accessed: 06/05/2025.

\bibitem{deepcorr}
Wei Wang, Ming Zhu, Xuewen Zeng, Xiaozhou Ye, and Yiqiang Sheng.
\newblock {Malware traffic classification using convolutional neural network for representation learning}.
\newblock In {\em International Conference on Information Networking (ICOIN)}, 2017.

\bibitem{xNIDS}
Feng Wei, Hongda Li, Ziming Zhao, and Hongxin Hu.
\newblock {{xNIDS}: Explaining Deep Learning-based Network Intrusion Detection Systems for Active Intrusion Responses}.
\newblock In {\em USENIX Security}, 2023.

\bibitem{bayesian-optimization}
Wikipedia.
\newblock {Bayesian Optimization}.
\newblock \url{https://en.wikipedia.org/wiki/Bayesian\_optimization}, last accessed: 06/05/2025.

\bibitem{winstein2013tcp}
Keith Winstein and Hari Balakrishnan.
\newblock {TCP ex machina: Computer-generated Congestion Control}.
\newblock In {\em ACM SIGCOMM Computer Communication Review (CCR)}, 2013.

\bibitem{gpu}
Wenji Wu and Phil Demar.
\newblock {A GPU-accelerated network traffic monitoring and analysis system}.
\newblock In {\em IEEE Conference on Computer Communications Workshops (INFOCOM WKSHPS)}, 2013.

\bibitem{mousika}
Guorui Xie, Qing Li, Yutao Dong, Guanglin Duan, Yong Jiang, and Jingpu Duan.
\newblock {Mousika: Enable General In-Network Intelligence in Programmable Switches by Knowledge Distillation}.
\newblock In {\em IEEE INFOCOM}, 2022.

\bibitem{rosetta}
Renjie Xie, Jiahao Cao, Enhuan Dong, Mingwei Xu, Kun Sun, Qi~Li, Licheng Shen, and Menghao Zhang.
\newblock {Rosetta: Enabling Robust {TLS} Encrypted Traffic Classification in Diverse Network Environments with {TCP-Aware} Traffic Augmentation}.
\newblock In {\em USENIX Security}, 2023.

\bibitem{xilinx-sn-1000}
Xilinx.
\newblock {Alveo SN1000 SmartNICs}.
\newblock \url{https://www.xilinx.com/content/dam/xilinx/publications/product-briefs/sn1000-product-brief.pdf}, last accessed: 06/05/2025.

\bibitem{xilinx-u250}
AMD Xilinx.
\newblock {Alveo U250 Data Center Accelerator Card}.
\newblock \url{https://www.xilinx.com/products/boards-and-kits/alveo/u250.html}, last accessed: 06/05/2025.

\bibitem{iisy-hotnets}
Zhaoqi Xiong and Noa Zilberman.
\newblock {Do Switches Dream of Machine Learning? Toward In-Network Classification}.
\newblock In {\em ACM HotNets}, 2019.

\bibitem{xsight-x2}
{Xsight Labs}.
\newblock {X2 Programmable Ethernet Switch}.
\newblock \url{https://xsightlabs.com/products/}, last accessed: 06/05/2025.

\bibitem{puffer}
Francis~Y. Yan, Hudson Ayers, Chenzhi Zhu, Sadjad Fouladi, James Hong, Keyi Zhang, Philip Levis, and Keith Winstein.
\newblock {Learning in situ: A Randomized Experiment in Video Streaming}.
\newblock In {\em USENIX NSDI}, 2020.

\bibitem{yan2018pantheon}
Francis~Y Yan, Jestin Ma, Greg~D Hill, Deepti Raghavan, Riad~S Wahby, Philip Levis, and Keith Winstein.
\newblock {Pantheon: The Training Ground for Internet Congestion-Control Research}.
\newblock In {\em USENIX ATC}, 2018.

\bibitem{bos}
Jinzhu Yan, Haotian Xu, Zhuotao Liu, Qi~Li, Ke~Xu, Mingwei Xu, and Jianping Wu.
\newblock {Brain-on-switch: towards advanced intelligent network data plane via NN-driven traffic analysis at line-speed}.
\newblock In {\em USENIX NSDI}, 2024.

\bibitem{planter}
Changgang Zheng, Mingyuan Zang, Xinpeng Hong, Liam Perreault, Riyad Bensoussane, Shay Vargaftik, Yaniv Ben-Itzhak, and Noa Zilberman.
\newblock {Planter: Rapid prototyping of in-network machine learning inference}.
\newblock {\em ACM SIGCOMM Computer Communication Review (CCR)}, 2024.

\bibitem{netbeacon}
Guangmeng Zhou, Zhuotao Liu, Chuanpu Fu, Qi~Li, and Ke~Xu.
\newblock {An Efficient Design of Intelligent Network Data Plane}.
\newblock In {\em USENIX Security}, 2023.

\end{thebibliography}


\label{totalpage}

\end{document}